\documentclass[12pt,a4paper]{article}
\usepackage{axodraw}
\usepackage{epsfig}

\advance\textheight by \topskip
\if@twoside \oddsidemargin -17pt \evensidemargin 00pt
\else \oddsidemargin 00pt \evensidemargin 00pt
\fi
\expandafter\ifx\csname mathrm\endcsname\relax\def\mathrm#1{{\rm #1}}\fi

\renewcommand{\theequation}{\thesection.\arabic{equation}}
\newcounter{saveeqn}

\makeatletter 
\@addtoreset{equation}{section}
\makeatother

\def\beq{\begin{equation}}
\def\eeq{\end{equation}}
\def\beqar{\begin{eqnarray}}
\def\eeqar{\end{eqnarray}}
\def\barr#1{\begin{array}{#1}}
\def\earr{\end{array}}
\def\bfi{\begin{figure}}
\def\efi{\end{figure}}
\def\btab{\begin{table}}
\def\etab{\end{table}}
\def\bce{\begin{center}}
\def\ece{\end{center}}

\def\nl{\nonumber\\}
\def\nln{\nonumber\\*[-1ex]\phantom{\fbox{\rule{0em}{2ex}}}}

\def\al{\alpha}
\def\be{\beta}
\def\ga{\gamma}
\def\de{\delta}

\def\la{\lambda}
\def\si{\sigma}

\def\De{\Delta}

\def\refeq#1{\mbox{(\ref{#1})}}
\def\reffi#1{\mbox{Fig.~\ref{#1}}}
\def\reffis#1{\mbox{Figs.~\ref{#1}}}
\def\refta#1{\mbox{Table~\ref{#1}}}

\def\refse#1{\mbox{Sect.~\ref{#1}}}

\def\refapp#1{\mbox{App.~\ref{#1}}}
\def\citere#1{\mbox{Ref.~\cite{#1}}}
\def\citeres#1{\mbox{Refs.~\cite{#1}}}

\def\solid{\raise.9mm\hbox{\protect\rule{1.1cm}{.2mm}}}
\def\dash{\raise.9mm\hbox{\protect\rule{2mm}{.2mm}}\hspace*{1mm}}
\def\dot{\rlap{$\cdot$}\hspace*{2mm}}

\def\solid{\raise.9mm\hbox{\protect\rule{12mm}{.2mm}}}
\def\dash{\raise.9mm\hbox{\protect\rule{1.6mm}{.2mm}}\hspace*{1mm}}
\def\dot{\raise.9mm\hbox{\protect\rule{0.8mm}{.2mm}}\hspace*{0.8mm}}
\def\dashdot{\raise.9mm\hbox{\protect\rule{.3mm}{.2mm}}\hspace*{.8mm}\raise.9mm\hbox{\protect\rule{1.3mm}{.2mm}}\hspace*{.8mm}}

\newcommand{\GeV}{\unskip\,\mathrm{GeV}}

\newcommand{\TeV}{\unskip\,\mathrm{TeV}}
\newcommand{\fba}{\unskip\,\mathrm{fb}}

\def\mathswitchr#1{\relax\ifmmode{\mathrm{#1}}\else$\mathrm{#1}$\fi}

\newcommand{\PW}{\mathswitchr W}
\newcommand{\PZ}{\mathswitchr Z}
\newcommand{\PA}{\mathswitchr A}

\newcommand{\PH}{\mathswitchr H}
\newcommand{\Pe}{\mathswitchr e}

\newcommand{\Pd}{\mathswitchr d}
\newcommand{\PD}{\mathswitchr D}

\newcommand{\Pu}{\mathswitchr u}
\newcommand{\PU}{\mathswitchr U}
\newcommand{\Ps}{\mathswitchr s}

\newcommand{\Pc}{\mathswitchr c}
\newcommand{\Pt}{\mathswitchr t}

\newcommand{\Pep}{\mathswitchr {e^+}}
\newcommand{\Pem}{\mathswitchr {e^-}}

\newcommand{\PWp}{\mathswitchr {W^+}}
\newcommand{\PWm}{\mathswitchr {W^-}}
\newcommand{\PWpm}{\mathswitchr {W^\pm}}
\newcommand{\Pp}{\mathswitchr {p}}

\newcommand{\PN}{\mathswitch {N}}
\newcommand{\PPi}{\mathswitchr i}

\def\mathswitch#1{\relax\ifmmode#1\else$#1$\fi}

\newcommand{\MW}{\mathswitch {M_\PW}}

\newcommand{\MZ}{\mathswitch {M_\PZ}}
\newcommand{\MH}{\mathswitch {M_\PH}}

\newcommand{\Mt}{\mathswitch {m_\Pt}}
\newcommand{\GW}{\mathswitch {\Gamma_\PW}}
\newcommand{\GZ}{\Gamma_{\PZ}}
\newcommand{\PL}{\mathswitch {P_\PL}}

\newcommand{\scrs}{\scriptscriptstyle}
\newcommand{\sw}{\mathswitch {s_{\scrs\PW}}}
\newcommand{\cw}{\mathswitch {c_{\scrs\PW}}}

\def\ie{i.e.\ }
\def\eg{e.g.\ }

\renewcommand{\O}{{\cal O}}
\newcommand{\Oa}{\mathswitch{{\cal{O}}(\alpha)}}

\newcommand{\SUtwo}{\mathrm{SU(2)}}

\newcommand{\CM}{\mathrm{CM}}

\newcommand{\Born}{\mathrm{Born}}

\newcommand{\rT}{{\mathrm{T}}}
\newcommand{\rL}{{\mathrm{L}}}

\newcommand{\rd}{{\mathrm{d}}}

\newcommand{\ri}{{\mathrm{i}}}

\newcommand{\M}{{\cal{M}}}

\newcommand{\virt}{\mathrm{virt}}

\newcommand{\LL}{\mathrm{LL}}

\newcommand{\TT}{\mathrm{TT}}

\newcommand{\LT}{\mathrm{LT}}
\newcommand{\TL}{\mathrm{TL}}

\newcommand{\elm}{{\mathrm{em}}}
\newcommand{\ew}{{\mathrm{ew}}}

\newcommand{\SC}{{\mathrm{LSC}}}
\renewcommand{\SS}{{\mathrm{SSC}}}
\newcommand{\cc}{{\mathrm{C}}}

\newcommand{\pre}{{\mathrm{PR}}}

\newcommand{\DPA}{{\mathrm{DPA}}}
\newcommand{\LPA}{{\mathrm{LPA}}}
\newcommand{\SPA}{{\mathrm{SPA}}}

\def\Re{\mathop{\mathrm{Re}}\nolimits}

\newcommand{\bew}{b^{\ew}}

\newcommand{\cew}{C^{\ew}}

\newcommand{\losmt}{\log{\left(\frac{\hat{s}}{\Mt^2}\right)}}
\newcommand{\los}{\log{\left(\frac{\hat{s}}{\MW^2}\right)}}
\newcommand{\Los}{\log^2{\left(\frac{\hat{s}}{\MW^2}\right)}}

\newcommand{\loZW}{\log{\left(\frac{\MZ^2}{\MW^2}\right)}}
\newcommand{\loWla}{\log{\left(\frac{\MW^2}{\la^2}\right)}}
\newcommand{\loWVa}{\log{\left(\frac{\MW^2}{M_{V^a}^2}\right)}}
\newcommand{\lots}{\log{\left(\frac{|\hat{t}|}{\hat{s}}\right)}}
\newcommand{\lous}{\log{\left(\frac{|\hat{u}|}{\hat{s}}\right)}}
\newcommand{\lotu}{\log{\left(\frac{|\hat{t}|}{|\hat{u}|}\right)}}

\newcommand{\lemphi}{l^\elm(m_{\varphi}^2)}

\newcommand{\lemW}{l^\elm(\MW^2)}

\newcommand{\Lemphi}{L^\elm(\hat{s},\lambda^2,m_\varphi^2)}

\newcommand{\PT}{P_{\mathrm{T}}}
\newcommand{\ET}{E_{\mathrm{T}}}
\newcommand{\PTmiss}{P_{\mathrm{T}}^{\mathrm{miss}}}
\newcommand{\PTcut}{P_{\mathrm{T}}^{\mathrm{cut}}}
\newcommand{\MT}{M_{\mathrm{T}}}
\newcommand{\Mcut}{M^{\mathrm{cut}}}

\makeatletter
\newcount\@tempcntc
\def\@citex[#1]#2{\if@filesw\immediate\write\@auxout{\string\citation{#2}}\fi
  \@tempcnta\z@\@tempcntb\m@ne\def\@citea{}\@cite{\@for\@citeb:=#2\do
    {\@ifundefined
       {b@\@citeb}{\@citeo\@tempcntb\m@ne\@citea
        \def\@citea{,\penalty\@m\ }{\bf ?}\@warning
       {Citation `\@citeb' on page \thepage \space undefined}}%
    {\setbox\z@\hbox{\global\@tempcntc0\csname
b@\@citeb\endcsname\relax}%
     \ifnum\@tempcntc=\z@ \@citeo\@tempcntb\m@ne
       \@citea\def\@citea{,\penalty\@m}
       \hbox{\csname b@\@citeb\endcsname}%
     \else
      \advance\@tempcntb\@ne
      \ifnum\@tempcntb=\@tempcntc
      \else\advance\@tempcntb\m@ne\@citeo
      \@tempcnta\@tempcntc\@tempcntb\@tempcntc\fi\fi}}\@citeo}{#1}}

\def\@citeo{\ifnum\@tempcnta>\@tempcntb\else\@citea
  \def\@citea{,\penalty\@m}%
  \ifnum\@tempcnta=\@tempcntb\the\@tempcnta\else
   {\advance\@tempcnta\@ne\ifnum\@tempcnta=\@tempcntb \else
\def\@citea{--}\fi
    \advance\@tempcnta\m@ne\the\@tempcnta\@citea\the\@tempcntb}\fi\fi}
\makeatother

\def\draftdate{\relax}
\def\mda{\relax}
\def\mua{\relax}
\def\mla{\relax}
\def\draft{
\def\thtystars{******************************}
\def\sixtystars{\thtystars\thtystars}
\typeout{}
\typeout{\sixtystars**}
\typeout{* Draft mode!
         For final version remove \protect\draft\space in source file *}
\typeout{\sixtystars**}
\typeout{}
\def\draftdate{\today}
\def\mua{\marginpar[\boldmath\hfil$\uparrow$]%
                   {\boldmath$\uparrow$\hfil}%
                    \typeout{marginpar: $\uparrow$}\ignorespaces}
\def\mda{\marginpar[\boldmath\hfil$\downarrow$]%
                   {\boldmath$\downarrow$\hfil}%
                    \typeout{marginpar: $\downarrow$}\ignorespaces}
\def\mla{\marginpar[\boldmath\hfil$\rightarrow$]%
                   {\boldmath$\leftarrow $\hfil}%
                    \typeout{marginpar: $\leftrightarrow$}\ignorespaces}
\def\Mua{\marginpar[\boldmath\hfil$\Uparrow$]%
                   {\boldmath$\Uparrow$\hfil}%
                    \typeout{marginpar: $\Uparrow$}\ignorespaces}
\def\Mda{\marginpar[\boldmath\hfil$\Downarrow$]%
                   {\boldmath$\Downarrow$\hfil}%
                    \typeout{marginpar: $\Downarrow$}\ignorespaces}
\def\Mla{\marginpar[\boldmath\hfil$\Rightarrow$]%
                   {\boldmath$\Leftarrow $\hfil}%
                    \typeout{marginpar: $\Leftrightarrow$}\ignorespaces}
\overfullrule 5pt
\oddsidemargin -15mm
\marginparwidth 29mm
}

\makeatletter

\def\eqnarray{\stepcounter{equation}\let\@currentlabel=\theequation
\global\@eqnswtrue
\global\@eqcnt\z@\tabskip\@centering\let\\=\@eqncr
$$\halign to \displaywidth\bgroup\hskip\@centering
  $\displaystyle\tabskip\z@{##}$\@eqnsel&\global\@eqcnt\@ne
  \hskip 2\arraycolsep \hfil${##}$\hfil
  &\global\@eqcnt\tw@ \hskip 2\arraycolsep $\displaystyle\tabskip\z@{##}$\hfil
   \tabskip\@centering&\llap{##}\tabskip\z@\cr}
\def\appendix{\par
 \setcounter{section}{0} \setcounter{subsection}{0}
 \def\thesection{\Alph{section}}}

\makeatother

\newcommand{\lsim}
{\;\raisebox{-.3em}{$\stackrel{\displaystyle <}{\sim}$}\;}
\newcommand{\gsim}
{\;\raisebox{-.3em}{$\stackrel{\displaystyle >}{\sim}$}\;}

\begin{document}
\tolerance=100000
\thispagestyle{empty}
\setcounter{page}{0}

\thispagestyle{empty}
\def\thefootnote{\fnsymbol{footnote}}
\setcounter{footnote}{1}
\null
\draftdate\hfill  PSI-PR-01-13
\\
\strut\hfill ZU-TH-35-01\\
\vskip 0cm
\vfill
\begin{center}
{\Large \bf
Electroweak-correction effects in gauge-boson pair production at the LHC
\par} \vskip 2.5em
{\large
{\sc E. Accomando$^1$, A.~Denner$^1$ and S. Pozzorini$^{1,2}$}}%
\\[.5cm]
$^1$ {\it Paul Scherrer Institut\\
CH-5232 Villigen PSI, Switzerland}
\\[0.3cm]
$^2$ {\it Institute of Theoretical Physics\\ University of Z\"urich, CH-8057 
Z\"urich, Switzerland}
\par
\end{center}\par
\vskip 2.0cm \vfill
{\bf Abstract:} \par
We have studied the effect of one-loop logarithmic electroweak
radiative corrections on $\PW\PZ$ and $\PW\ga$ production processes at
the LHC.  We present analytical results for the leading-logarithmic
electroweak corrections to the corresponding partonic processes
$\bar\Pd\Pu\to\PW\PZ,\PW\ga$. Using the leading-pole approximation we
implement these corrections into Monte Carlo programs for $\Pp\Pp\to
l\nu_ll^\prime\bar{l^\prime}, l\nu_l\gamma$.  We find that electroweak
corrections lower the predictions by 5--20\% in the physically
interesting region of large transverse momentum and small rapidity
separation of the gauge bosons.
\par
\vskip 1cm
\noindent
October 2001
\par
\null
\setcounter{page}{0}
\clearpage
\def\thefootnote{\arabic{footnote}}
\setcounter{footnote}{0}

\section{ Introduction }

Vector-boson pair production provides us with an important testing
ground for the non-abelian structure of the Standard Model (SM). While
gauge-boson properties, such as masses and couplings to fermions, have
already been measured with great accuracy at LEP and Tevatron,
vector-boson self-interactions have not been tested with comparable
precision.  New physics occurring at energy scales much larger than
those probed directly at forthcoming experiments could modify the
structure of these interactions.  These modifications are parametrized
in terms of anomalous couplings in the Yang--Mills vertices.

In the last few years, the contribution of trilinear gauge-boson
couplings was directly measured via vector-boson pair production at
LEP2 and Tevatron. While, in particular, LEP2 has been able to produce
$\PWp\PWm$ pairs with high statistics, nevertheless all these events
were generated at rather modest centre-of-mass (CM) energies
($E_{\mathrm{CM}}\lsim210\GeV$). The effect of anomalous couplings is,
on the other hand, expected to be strongly enhanced by increasing the
invariant mass of the gauge-boson pair $M_{VV'}$
($V,V'=\PW,\PZ,\gamma$), since these couplings in general spoil the
unitarity cancellations for longitudinal gauge bosons. Hence, at
future colliders it will be useful to analyse the di-boson production
at the highest possible CM energies.

Moreover, vector-boson pairs constitute a background to other kinds of
new-physics searches. One of the gold-plated signals for supersymmetry
at hadron colliders is chargino--neutralino pair production, which
would give rise to final states with three charged leptons and missing
transverse momentum \cite{Matchev:1999nb}; the primary background to
this signature is given by $\PW\PZ$ or $\PW\gamma^*$ production.
Leptonic final states, coming from $\PW\PZ$ or $\PW\gamma^*$, could
also fake $\PW\PZ$ vector-boson scattering signals, or $\PW^\pm
\PW^\mp$ and $\PW^\pm \PW^\pm$ scattering signals if one of the
charged leptons is lost in the beam pipe, which are again expected to
be enhanced at high CM energies \cite{Bagger:1994zf}.

In the next years, hadron colliders will be the main source of
vector-boson pairs with large invariant masses $M_{VV^\prime}$.
Tevatron Run II will collect from tens to hundreds of events,
depending on the particular process.  The large Hadron Collider (LHC)
will further increase the event number by roughly two orders of
magnitude \cite{lhcrep}.  Owing to the expected increase in
statistics, theoretical predictions must reach high accuracy to allow
for a decent analysis of the data.

Hadronic di-boson production has received a lot of attention (for a
review on the subject see \citere{lhcrep}).  Originally computed by
treating $\PW$ and $\PZ$ bosons as stable particles, tree-level cross
sections for $\PW^+\PW^-$, $\PW^\pm \PZ$, $\PZ\PZ$, $\PW^\pm \gamma$,
and $\PZ \gamma$ production and decay have been updated by evaluating,
in narrow-width approximation but retaining spin information via
decay-angle correlations, the doubly-resonant contribution to the
four-fermion final states (e.g. $q\bar q^\prime\to \PW^\pm \PZ \to
4f$) \cite{Dixon:1999di} and the resonant contribution to two-fermion
plus photon final states (e.g.  $q\bar q^\prime\to \PW^\pm \gamma \to
2f+\gamma$) \cite{DeFlorian:2000sg}. As a further step, in the last
few years Monte Carlo programs \cite{mc} have included the full $q\bar
q^\prime\to 4f$ amplitude, by taking into account finite-width effects
and the irreducible background owing to non-doubly-resonant diagrams.

The $\O(\alpha_s)$ QCD corrections to gauge-boson pair-production and
decay have been extensively analysed by many authors.  Gauge-boson
pair-production cross sections have been calculated at next-to-leading
order (NLO) accuracy retaining the full spin correlations of the
leptonic decay products. Several NLO Monte Carlo programs have been
implemented and cross checked so that complete $\O(\alpha_s)$
corrections are now available \cite{Dixon:1999di,DeFlorian:2000sg,mc}.
QCD corrections turn out to be quite significant at LHC energies.
They can increase the lowest-order cross section by a factor two if no
cuts are applied and by one order of magnitude for large transverse
momentum or large invariant mass of the vector bosons
\cite{Ohnemus:1991gb,Frixione:1992pj}.  By including a jet veto, their
effects can be drastically reduced to the order of tens of per cent
\cite{Baur:1995aj,Dixon:1999di}, but in any case they have to be
considered to get realistic and reliable estimates of total cross
sections and distributions.

In view of the envisaged precision of a few per cent at the LHC, also
a discussion of electroweak corrections is in order.  For single
$\PW$- and $\PZ$-boson production, $\O(\alpha )$ corrections have been
computed taking into account the full QED and weak contributions
\cite{Baur:1999kt}.  For gauge-boson pair production at hadron
colliders, the electroweak corrections have been taken into account
only via an effective mixing angle.

As well known, the impact of $\O(\alpha )$ electroweak contributions
grows with increasing energy.  Analyses of the high-energy behaviour
of electroweak corrections in general and for specific $\Pep\Pem$ and
$\gamma\gamma$ processes have already been performed revealing effects
which should be clearly visible at future linear colliders (see for
instance \citeres{Beenakker:1993tt,ewee}).  At high energies the
electroweak corrections are dominated by double and single logarithms
of the ratio of the energy to the electroweak scale. In
\citeres{Denner:2001jv,Denner:2001gw} it has been shown that the
leading-logarithmic one-loop corrections to arbitrary electroweak
processes factorize into the tree-level amplitudes times universal
correction factors. These results represent a process-independent
recipe for the calculation of leading logarithmic corrections.

Using the method of \citeres{Denner:2001jv,Denner:2001gw}, we
investigate in this paper the effect of leading-logarithmic
electroweak corrections to the hadronic production of $\PW^\pm \PZ$
and $\PW^\pm\gamma$ pairs in the large-invariant-mass region of the
hard process at the LHC.  Since the aim of this paper is to describe
the structure of the $\O(\alpha )$ electroweak corrections and to give
an estimate of their size, we have not included QCD corrections.
Also, QED corrections are not fully considered as they strictly depend
on the experimental setup.  We omit all infrared-singular terms
originating from the massless photon, as explained in
\refapp{app:one-loopcorr} and focus on the contributions of the
leading electroweak logarithms originating from above the electroweak
scale.

The simplest experimental analyses of gauge-boson pair production will
rely on purely leptonic final states. Semi-leptonic channels, where
one of the vector bosons decays hadronically, have been analysed at
the Tevatron \cite{semitev} showing that these events suffer from the
background due to the production of one vector boson plus jets via
gluon exchange.  For this reason, we choose to analyse only di-boson
production where both gauge bosons decay leptonically into $e$ or
$\mu$.
  
The paper is organized as follows: in \refse{se:processes} we give
some details on the general setup of our calculation. In
\refse{se:ewrc} and \refapp{app:corr} the logarithmic electroweak
one-loop corrections are examined and presented in analytical form.
Section \ref{se:WZ} contains a numerical discussion for $\PW\PZ$
production and decay, while \refse{se:WA} covers $\PW\gamma$
production and decay.  Our findings are summarized in
\refse{se:concl}.

\section{Processes and their computation}
\label{se:processes}

We consider in detail two classes of processes,
\renewcommand{\labelenumi}{(\roman{enumi})}
\begin{enumerate}
\item $\Pp\Pp\to l\nu_ll^\prime\bar{l^\prime}$, with $l,l^\prime =\Pe,\mu$,
\qquad and
\item $\Pp\Pp\to l\nu_l\gamma$, with $l=\Pe,\mu$.
\end{enumerate}
The first class is characterized by three isolated charged leptons
plus missing energy in the final state.  In our notation, $l\nu_l$
indicates both $l^-\bar\nu_l$ and $l^+\nu_l$. This kind of processes
includes $\PW\PZ$ production as intermediate state. The second class
is instead related to $\PW\gamma$ production.
Both classes of processes are described by the formula
\beqar
\rd\si^{h_1 h_2}(P_1,P_2,p_f) = \sum_{i,j}\int\rd x_1 \rd x_2~
f_{i,h_1}(x_1,Q^2)f_{j,h_2}(x_2,Q^2) \,\rd\hat\si^{ij}(x_1P_1,x_2P_2,p_f),
\eeqar
where $p_f$ summarizes the final-state momenta, $f_{i,h_1}$ and
$f_{j,h_2}$ are the distribution functions of the partons $i$ and $j$ in
the incoming hadrons $h_1$ and $h_2$ with momenta $P_1$ and $P_2$,
respectively, $Q$ is the factorization scale, and $\hat\si^{ij}$ represent the 
cross sections for the
partonic processes. Since the two incoming hadrons are protons and 
we sum over final states with opposite charges, we find
\beqar\label{eq:convol}
\rd\si^{h_1h_2}(P_1,P_2,p_f) = \int\rd x_1 \rd x_2 &&\sum_{U=u,c}\sum_{D=d,s}
\Bigl[f_{\bar\PD,\Pp}(x_1,Q^2)f_{\PU,\Pp}(x_2,Q^2)\,\rd\hat\si^{\bar\PD\PU}
(x_1P_1,x_2P_2,p_f)
\nl&&{}
+f_{\bar\PU,\Pp}(x_1,Q^2)f_{\PD,\Pp}(x_2,Q^2)\,\rd\hat\si^{\bar\PU\PD}
(x_1P_1,x_2P_2,p_f)
\nl&&{}
+f_{\PD,\Pp}(x_1,Q^2)f_{\bar\PU,\Pp}(x_2,Q^2)\,\rd\hat\si^{\PD\bar\PU}
(x_1P_1,x_2P_2,p_f)
\nl&&{}
+f_{\PU,\Pp}(x_1,Q^2)f_{\bar\PD,\Pp}(x_2,Q^2)\,\rd\hat\si^{\PU\bar\PD}
(x_1P_1,x_2P_2,p_f)\Bigr]
\eeqar
in leading order of QCD.

The tree-level amplitudes for the partonic processes have been
generated by means of PHACT \cite{phact}, a set of routines based on
the helicity-amplitude formalism of \citere{helamp}. For the numerical
results presented here, we have used the fixed-width scheme with $\GZ
=2.512\GeV$ and $\GW =2.105\GeV$, and the input masses
$\MZ=91.187\GeV$ and $\MW=80.45\GeV$. The weak mixing angle is fixed
by $\sw^2=1-\MW^2/\MZ^2$.  Moreover, we adopted the so called
$G_{\mu}$-scheme, which effectively includes higher-order
contributions associated with the running of the electromagnetic
coupling and the leading universal two-loop $\Mt$-dependent
corrections. This corresponds to parametrize the lowest-order matrix
element in terms of the effective coupling
$\alpha_{G_{\mu}}=\sqrt{2}G_{\mu}\MW^2\sw^2/\pi$. However, we use
$\alpha(0)=1/137.036$ for the coupling of the real photon in
$\Pp\Pp\to l\nu_l\ga$, \ie we multiply the corresponding cross
sections in the $G_{\mu}$-scheme by $\al(0)/\alpha_{G_{\mu}}$.
Additional input parameters are the quark-mixing matrix elements whose
values have been taken to be $|V_{\Pu\Pd}|=|V_{\Pc\Ps}|=0.975$,
$|V_{\Pu\Ps}|=|V_{\Pc\Pd}|=0.222$, and zero for all other relevant
matrix elements.

As to parton distributions, we have used CTEQ(5M1) \cite{cteq} at the 
factorization scales
\begin{equation}
Q^2={1\over 2}\left (\MW^2+\MZ^2+\PT^2(l\nu_l)+\PT^2(l^\prime
\bar{l^\prime})\right )
\end{equation}
and
\begin{equation}
Q^2={1\over 2}\left (\MW^2+\PT^2(l\nu_l)+\PT^2(\gamma )\right )
\end{equation}
for $\PW\PZ$ and $\PW\gamma$ production processes, respectively, where
$\PT$ denotes the transverse momentum. This scale choice appears to be
appropriate for the calculation of differential cross sections, in
particular for vector-boson transverse-momentum distributions
\cite{Frixione:1992pj,Dixon:1999di}.

We have, moreover, implemented a general set of cuts, proper for LHC 
analyses, defined as follows:
\begin{itemize}
\item {lepton transverse momentum $\PT(l)>20\GeV$},
  
\item {missing transverse momentum $\PTmiss> 20\ (50)\GeV$ for
    $\PW\PZ$ ($\PW\gamma$)},
  
\item {lepton pseudo-rapidity $|\eta_l |< 3\ (2.5)$ for $\PW\PZ$
    ($\PW\gamma$)}, where $\eta_i=-\log\left (\tan\theta_i/2\right )$,
  $\theta_i$ is the polar angle of particle $i$
with respect to the beam, and $i=l,\gamma$,
\item {rapidity--azimuthal-angle separation $\Delta R_{l\gamma}=
    \sqrt{(\eta_l-\eta_\gamma )^2+(\phi_l -\phi_\ga)^2}>0.7$ between
    charged lepton and photon for $\PW\ga$.}
\end{itemize}
For the different processes considered, we have also used further 
cuts which are described in due time. In the following sections, we present 
results for the LHC at $\CM$ energy $\sqrt s=14\TeV$ and an integrated 
luminosity $L=100\fba^{-1}$.

\section {Electroweak \boldmath $\O(\alpha$) corrections}
\label{se:ewrc}

We are interested in the electroweak $\O(\alpha$) corrections to the
processes $\Pp\Pp\to l\nu_ll^\prime\bar{l^\prime}$ and $\Pp\Pp\to
l\nu_l\ga$ in the region of phase space where these are dominated by
the gauge-boson pair-production subprocesses $\Pp\Pp\to\PW\PZ$ and
$\Pp\Pp\to\PW\ga$, respectively. In this region, the dominant
contributions are those that are enhanced by the resonant propagators
of the $\PW$ boson and in the first process also of the $\PZ$ boson.
These can be most effectively calculated in the so-called leading-pole
approximation (LPA), which is a double-pole approximation (DPA) for
$\Pp\Pp\to \PW\PZ\to l\nu_ll^\prime\bar{l^\prime}$ and a single-pole
approximation (SPA) for $\Pp\Pp\to \PW\ga\to l\nu_l\ga$. The LPA has
been successfully applied for the calculation of electroweak
corrections to \PW-pair production
\cite{Beenakker:1999gr,Denner:2000bj,Jadach:2000kw}.

At tree level, the DPA for the partonic process $qq'\to \PW\PZ\to
l\nu_ll^\prime\bar{l^\prime}$ reads
\newcommand{\laW}{\la}
\newcommand{\laZ}{\la'}
\newcommand{\laA}{\la'}
\beqar\label{eq:BornWZ}
\M_{\Born,\DPA}^{qq'\to\PW\PZ\to l\nu_ll^\prime\bar{l^\prime}} &=&
\frac{\PPi}{p_\PW^2-\MW^2+\ri\MW\GW}\;\frac{\PPi}{p_\PZ^2-\MZ^2+\ri\MZ\GZ} \nl
&&
\times\sum_{\laW,\laZ}\M_{\Born}^{qq'\to\PW_{\laW}\PZ_{\laZ}}
\M_\Born^{\PW_{\laW}\to l\nu_l}\M_\Born^{\PZ_{\laZ}\to l'\bar{l'}},
\eeqar
where the gauge-dependent doubly-resonant contribution is replaced by
the well-defined gauge-independent residue, and
$\M_{\Born}^{qq'\to\PW_{\laW}\PZ_{\laZ}}$, $\M_\Born^{\PW_{\laW}\to
  l\nu_l}$, and $\M_\Born^{\PZ_{\laZ}\to l'\bar{l'}}$ denote the
on-shell Born matrix elements for the boson production and decay
processes.  The sum runs over the physical helicities
$\laW,\laZ=0,\pm1$ of the on-shell projected \PW~and \PZ~bosons (see
App.~A of \citere{Denner:2000bj} for details), while the momenta of
the virtual $\PW$ and $\PZ$ bosons are denoted by $p_\PW$ and $p_\PZ$,
respectively. For the process $qq'\to \PW\ga\to l\nu_l\ga$, the SPA is
given by
\beqar\label{eq:BornWA}
\M_{\Born,\SPA}^{qq'\to\PW\ga_{\laA}\to l\nu_l\ga_{\laA}} &=&
\frac{\PPi}{p_\PW^2-\MW^2+\ri\MW\GW}
\sum_{\laW}\M_{\Born}^{qq'\to\PW_{\laW}\ga_{\laA}}
\M_{\Born}^{\PW_{\laW}\to l\nu_l}.
\eeqar

In LPA, the $\O(\alpha)$ electroweak corrections to boson-production
processes can be divided into two classes, factorizable and
non-factorizable corrections. The non-factorizable corrections, \ie
those contributions that cannot be associated with either boson
production or decay, have been evaluated for boson-pair production in
$\Pep\Pem$ annihilation in \citeres{Beenakker:1997bp,Denner:1998ia}.
There, these corrections turned out to be
small. We assume that this holds as well for the similar processes considered
here and do not consider non-factorizable corrections any further.

Moreover, we do not include real corrections and restrict our
discussion to the infrared-finite part of the virtual factorizable
corrections, as defined in \refapp{app:one-loopcorr}. These
contributions can be expressed in terms of the corrections to the
on-shell boson production and decay subprocesses. The matrix element
for the virtual corrections to the process $qq'\to\PW\PZ\to
l\nu_ll^\prime\bar{l^\prime}$ can be written as
\beqar\label{eq:corrWZ}
\de\M_{\virt,\DPA}^{qq'\to\PW\PZ\to l\nu_ll^\prime\bar{l^\prime}} &=&
\frac{\PPi}{p_\PW^2-\MW^2+\ri\MW\GW}\;\frac{\PPi}{p_\PZ^2-\MZ^2+\ri\MZ\GZ} \nl
&&
\sum_{\laW,\laZ}\biggl\{\de\M_{\virt}^{qq'\to\PW_{\laW}\PZ_{\laZ}}
\M_{\Born}^{\PW_{\laW}\to l\nu_l}\M_{\Born}^{\PZ_{\laZ}\to l'\bar{l'}}\nl
&&{}+\M_{\Born}^{qq'\to\PW_{\laW}\PZ_{\laZ}}
\de\M_{\virt}^{\PW_{\laW}\to l\nu_l}\M_{\Born}^{\PZ_{\laZ}\to l'\bar{l'}}\nl
&&{}+\M_{\Born}^{qq'\to\PW_{\laW}\PZ_{\laZ}}
\M_{\Born}^{\PW_{\laW}\to l\nu_l}\de\M_{\virt}^{\PZ_{\laZ}\to l'\bar{l'}}\biggr\},
\eeqar
where $\delta\M_{\virt}^{qq'\to\PW_{\laW}\PZ_{\laZ}}$,
$\delta\M_\virt^{\PW_{\laW}\to l\nu_l}$, and
$\delta\M_\virt^{\PZ_{\laZ}\to l'\bar{l'}}$ denote the virtual
corrections to the on-shell matrix elements for the boson production
and decay processes.  A similar expression holds for the process
$qq'\to\PW\ga\to l\nu_l\ga$.

We focus in particular on the corrections involving single and double
enhanced electroweak logarithms at high energies, \ie on $\O(\alpha )$
contributions proportional to $\al\log^2(\hat{s}/\MW^2)$ or
$\al\log(\hat{s}/\MW^2)$, where $\sqrt{\hat{s}}$ is the CM energy of
the partonic subprocess.  The logarithmic approximation yields the
dominant corrections at CM energies large compared to the gauge-boson
masses, $\hat{s}\gg\MW^2$.  In the high-energy limit, however, there
might be also enhanced non-logarithmic contributions that are a priori
relevant. In general they contain constant terms proportional to
$\MH^2/\MW^2$ and $\Mt^2/\MW^2$. While for transverse gauge bosons,
these Higgs- and top-mass-dependent corrections are entirely due to
renormalization effects and can be effectively accounted for by using
the $G_\mu$-scheme, for longitudinal gauge bosons additional
contributions of this kind exist.  These non-logarithmic $\O(\alpha )$
contributions are process-dependent. For $\Pep\Pem\to\PWp\PWm$, where
complete $\O(\alpha )$ corrections and their high-energy limit are
available \cite{Beenakker:1993tt}, the above terms turn out to be of
order of a few per cent. We can then qualitatively assume that this
holds as well for similar processes like hadronic di-boson production,
even if only an exact computation could really furnish a precise
statement on this point.  Neglecting non-logarithmic terms can
therefore be considered a reasonable approximation at the LHC, where
the experimental accuracy in the high-energy regime is at the
few-per-cent level.

Since the decay processes involve no large-energy variable, the
corresponding virtual corrections vanish in the logarithmic
approximation. As a consequence, we do not consider in the following
the last two contributions on the right-hand side of
\refeq{eq:corrWZ}. Moreover, for the boson production processes
$qq'\to\PW\PZ,\PW\ga$ we take into account only the correction to the
dominating channels involving two transverse ($\TT$) or two
longitudinal ($\LL$) gauge bosons. The contributions of the mixed
($\LT$, $\TL$) channels are suppressed relatively to the others by
factors of $\MW/\sqrt{\hat{s}}$ in the high-energy limit (see
\reffi{fi:WZ_born_ecm}), and thus the corresponding corrections are
unimportant.

The logarithmic virtual electroweak corrections to the dominating
channels are calculated using the general method given in
\citeres{Denner:2001jv,Denner:2001gw}. The corresponding analytical
expressions for the processes $\bar\Pd\Pu\to\PW^+_\la N_\la$,
$N=\PZ,\ga$, are given in \refapp{app:corr}. Those for
$\Pd\bar\Pu\to\PW^-_\la N_\la$ are derived via CP symmetry [see
\refeq{CPsymm}].  Our predictions are obtained by considering the
matrix element squared
\begin{equation}\label{M2ddef}
|\M|^2=|\M_{\Born}|^2+2\Re\left [\M_{\Born,\LPA}\delta \M_{\virt,\LPA}^{\dagger}\right ],
\end{equation}
where $\M_{\Born}$ is the exact Born amplitude, while the $\O(\alpha)$
contribution is computed in LPA based on \refeq{eq:BornWZ}--%
\refeq{eq:corrWZ}, and the formulas given in \refapp{app:corr}. In the
high-energy limit, a reasonable approach is to neglect fermion and
boson masses, as compared with $\sqrt{\hat s}$, wherever possible. The
expressions given in \refapp{app:corr} are based on this
approximation.  However, we take into account the exact kinematics by
evaluating the complete four-fermion or two-fermion-plus-photon phase
space and use the exact values of the kinematical invariants in all
formulas.  Moreover, we do not use the high-energy approximations
\refeq{longbornWN} and \refeq{trabornWN} in the correction factors but
we implement the $\O(\alpha )$ contributions according to the full
expressions given in \refeq{SCWZcorr}, \refeq{duSCgen},
\refeq{SSCcorrections}, and \refeq{CcorrectionsL}--\refeq{eq:deCPRTT}
with the exact (SU(2)-transformed) Born matrix elements.  Owing to our
choice of the input-parameter scheme, the terms proportional to
$\De\al(\MW^2)$ in \refeq{RGWW} and \refeq{eq:deCPRTT} are omitted,
since these are already taken into account by using $\al_{G_\mu}$
instead of $\al(0)$ as input.

In the universal logarithmic corrections given in \refapp{app:corr},
the pure angular-dependent logarithms, such as
$\alpha\log^2(|\hat{r}|/\hat{s})$ and $\alpha\log (|\hat{r}|/\hat{s})$
with $\hat{r}$ equal to the Mandelstam variables $\hat{t}$ and
$\hat{u}$ of the partonic subprocess, are not included.  The validity
of this approximation relies therefore on the assumption that all the
variables $\hat{s}$, $|\hat{t}|$, and $|\hat{u}|$ are large compared
with $\MW^2$ and approximately of the same size,
\beq\label{HEA}
\hat{s}\sim |\hat{t}|\sim |\hat{u}|\gg \MW^2.
\eeq
This implies that the produced gauge bosons have to be emitted at
sufficiently large angles with respect to the beam. Hence, the
validity range of the high-energy logarithmic approximation for the
radiative corrections corresponds to the central region of the boson
scattering angle in the di-boson rest-frame. For $s$-channel
processes, integrating over the full angular domain does not affect
the reliability of the result at logarithmic level, since the
neglected pure angular-dependent logarithms would give rise only to
subleading constant terms, if included. For $t$-channel dominated
scatterings like $\PW\PZ$ or $\PW\ga$ production, the situation is
instead more delicate. The $t$-channel pole in the Born matrix element
gives rise to additional enhanced logarithms when integrated over the
full kinematical range.  Since these terms are not included in our
$\O(\alpha )$ analysis, we have to take care that we do not get
sizeable contributions from small scattering angles with respect to
the beam. On the other hand, our formulas do not fake spurious
contributions as long as $\hat{s},|\hat{t}|,|\hat{u}|\not\ll\MW^2$,
since the large logarithms become small for
$\hat{s},|\hat{t}|,|\hat{u}|\sim\MW^2$.

\section {\boldmath$\PW^\pm\PZ$ production}
\label{se:WZ}

In this section, we present some cross sections and distributions for
the leptonic processes $\Pp\Pp\to l\nu_ll^\prime\bar{l^\prime}$ with
$l,l^\prime =\Pe,\mu$.  These final states allow to analyse $\PW\PZ$
production and thus in particular to test trilinear gauge-boson
couplings. Systematic studies of the effect of anomalous couplings on
the hadronic production of gauge-boson pairs have shown that
deviations from the SM cross sections should be particularly enhanced
when gauge bosons are produced at high CM energies and at large
scattering angles in the di-boson rest-frame.  The same kinematical
region is also proper to search for scatterings of strongly
interacting vector bosons.

It is therefore particularly interesting to study the electroweak
corrections in these kinematical configurations, where their effect is
also expected to be more sizeable.  As an illustration of the
behaviour and the size of the $\O(\alpha )$ contributions, we have
chosen to analyse the distribution of the reconstructed $\PZ$-boson
transverse momentum $\PT(l^\prime\bar{l^\prime})$.  The $\PT$ variable
is commonly used at hadron colliders because large $\PT$ requires high
CM energies and large angles.  We study also pure angular observables of
interest in the high-energy regime of the hard scattering.

\subsection{Born level}
\label{se:WZ:Born}

\begin{figure}
  \begin{center}
\epsfig{file=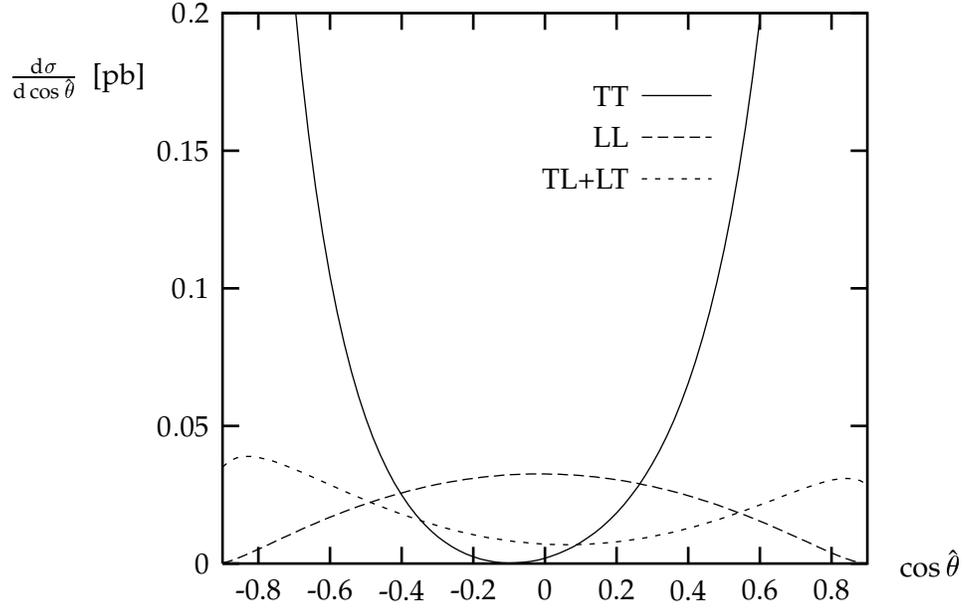, angle=0}
\end{center}
\caption{Lowest-order angular distributions for the process
        $\bar\Pd\Pu\rightarrow\PW^+_\lambda\PZ_{\lambda^\prime}$ at
        $E_{\CM}=500\GeV$.  Here $\lambda ,\lambda^\prime$ denote the
        transverse (T) or longitudinal (L) helicities.}
\label{fi:WZ_born_theta}
\end{figure}

\begin{figure}
  \begin{center}
\epsfig{file=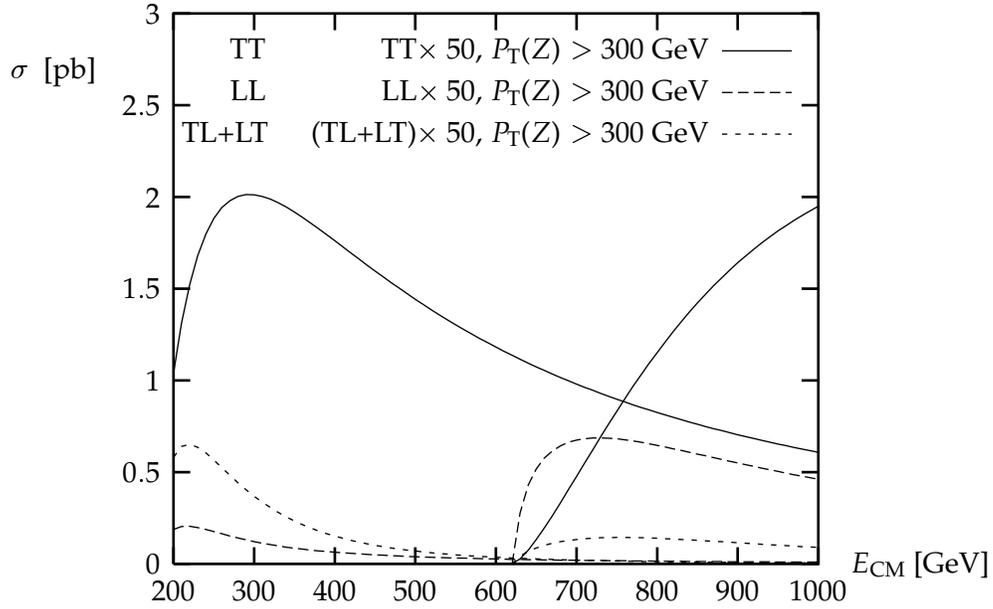, angle=0}
\end{center}  
\caption{Born cross sections for the process 
  $\bar\Pd\Pu\rightarrow\PW^+_\lambda\PZ_{\lambda^\prime}$ as a
  function of $E_{\CM}$ with $\lambda ,\lambda^\prime$ as in
  \reffi{fi:WZ_born_theta}. From left to right, the three legends
  refer to the left-side curves and to the right-side ones respectively,
  as explained in the text.}
\label{fi:WZ_born_ecm}
\end{figure}

\newcommand{\guL}{g_{\Pu,\rL}}%
\newcommand{\gdL}{g_{\Pd,\rL}}%
We start recalling basic properties of the Born amplitude, which are
useful later in discussing radiative-correction effects. In
\reffi{fi:WZ_born_theta}, just for explicative purposes, we have
plotted the on-shell Born cross section for the partonic process $\bar
\Pd\Pu\to \PW^+\PZ$ as a function of the angle $\hat{\theta}$ between
the $\bar \Pd$~quark and the $\PZ$~boson in the $\bar \Pd\Pu$ CM
frame, at fixed energy $E_{\CM}=500\GeV$ and before any convolution
with quark distribution functions.
We have reported the different helicity contributions separately. As
can be seen, the transverse component $\sigma_{\TT}$ shows the
well-known radiation zero for
$\cos{\hat{\theta}}=(\guL+\gdL)/(\guL-\gdL)=-\sw^2/(3\cw^2)\approx-0.1$
\cite{Baur:1994prl,Baur:1995aj}, where $\guL$ and $\gdL$ represent the 
$\PZ$-boson couplings to left-handed up and down quarks, respectively, and is
strongly peaked in the forward and backward directions. The
longitudinal contribution $\sigma_{\LL}$ is instead concentrated in
the central region, at large angle of the $\PZ$-boson with respect to
the incoming quarks. Integrating over the angle from $0^\circ$ to
$180^\circ$, one obtains the total cross sections shown in
\reffi{fi:WZ_born_ecm} (see the three curves on the left side) as a
function of the energy.
As expected, the dominant contribution is given by $\sigma_{\TT}$, and above 
$300\GeV$ all polarized cross sections decrease with energy.

The behaviour of the polarized cross sections depends, however, on the
selected kinematical region. If we consider the region of phase space
characterized by a large transverse momentum of the $\PZ$ boson, the
relative size of the different helicity components and the shape of
the curves change sensibly.  As before, we plot the cross sections
versus the CM energy but now for $\PT(\PZ)>300\GeV$ (see the three curves
starting at around $600\GeV$ in \reffi{fi:WZ_born_ecm}). In this case the $\LL$
contribution dominates at smaller energies, while the $\TT$ component
increases with energy and takes over at high CM energies. This is due to the 
fact that the above-mentioned $\PT$ cut translates into a minimum CM energy,
$E_{\CM}\simeq 624\GeV$, and limits the allowed range of the
scattering angle of the $\PZ$~boson. Hence, at low energies, the
allowed angular region is strictly central and the $\LL$ component dominates. 
At larger energies, the allowed kinematical range increases by including
smaller angles, and the $\TT$ contribution rapidly grows, soon
overwhelming the $\LL$ part.

\begin{figure}
  \begin{center}
\epsfig{file=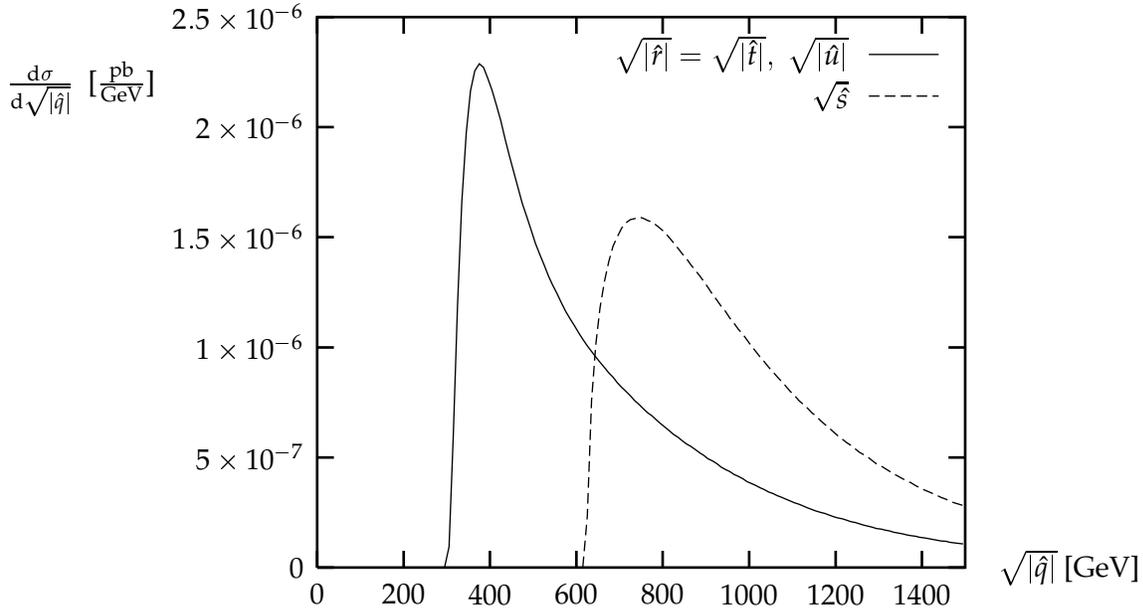, angle=0}
\end{center}
\caption{Lowest-order distributions in the invariants $\sqrt{\hat
    s}$ and $\sqrt{|\hat{r}|}$ as defined in the text for the full
  process $\Pp\Pp\rightarrow l\nu_ll^\prime\bar{l^\prime}$ at
  $\sqrt{s}=14\TeV$. Standard cuts and
  $\PT(l^\prime\bar{l^\prime})>300\GeV$ are applied.}
\label{fi:WZ_born_inv}
\end{figure}

Of course, one has to consider the additional effect due to the partonic
distribution functions, which in turn decrease with increasing momentum fractions $x_i$ and therefore with increasing CM energy 
$\sqrt{\hat{s}}=\sqrt{x_1x_2s}$. The net
result is shown in \reffi{fi:WZ_born_inv}, where the distribution in the
hard-scattering energy $\sqrt{\hat s}$ is plotted. 
Here and in the following we consider the full process $\Pp\Pp\to 4f$,
summed over all electron and muon final states, 
$\Pe^\pm{\mathop{\nu}\limits^{\scriptscriptstyle(-)}}_{\!\!\Pe}\Pep\Pem$,
$\Pe^\pm{\mathop{\nu}\limits^{\scriptscriptstyle(-)}}_{\!\!\Pe}\mu^+\mu^-$,
$\mu^\pm{\mathop{\nu}\limits^{\scriptscriptstyle(-)}}_{\!\!\mu}\Pep\Pem$,
and
$\mu^\pm{\mathop{\nu}\limits^{\scriptscriptstyle(-)}}_{\!\!\mu}\mu^+\mu^-$.
We have moreover applied our standard cuts as defined in
\refse{se:processes} and the additional cut $\PT(l'\bar l')>300\GeV$ on
the reconstructed $\PZ$~boson. As can be seen, despite the suppression
resulting from the decrease of the parton distributions with energy,
roughly 50$\%$ of the contribution to the total cross section comes
from the high-energy region $\sqrt{\hat s}>1\TeV$.  We come back to
this point later when discussing radiative corrections.

\begin{figure}
  \begin{center}
\epsfig{file=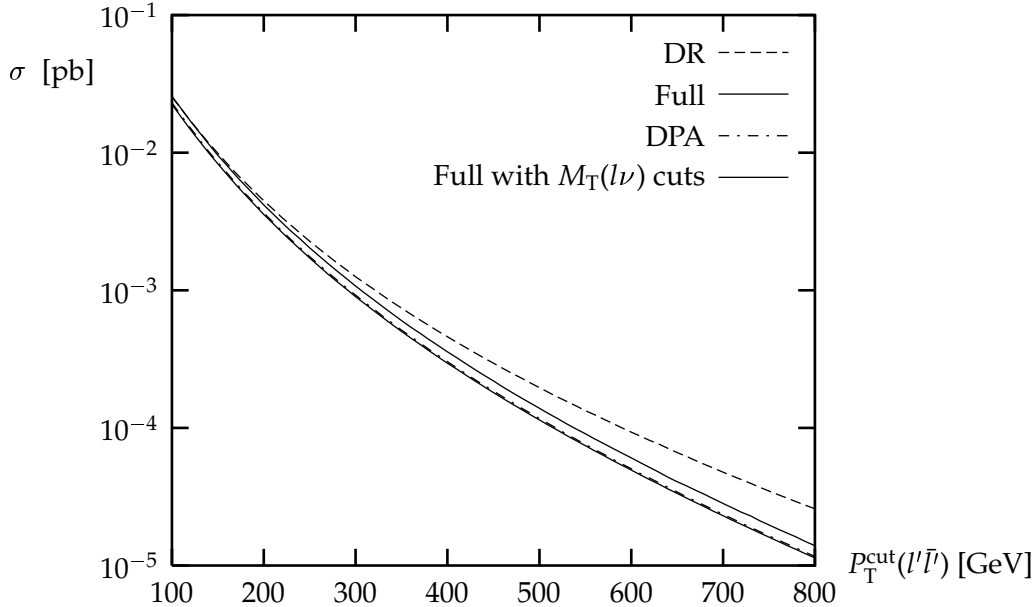, angle=0}
\end{center}  
\caption{Born cross section for the full process 
  $\Pp\Pp\to l\nu_ll^\prime\bar{l^\prime}$ at $\sqrt{s}=14\TeV$ as a
  function of the cut on the transverse momentum of the reconstructed
  $\PZ$~boson. Standard cuts are applied.}
\label{fi:WZ_born_pt}
\end{figure}

As explained in \refse{se:ewrc}, the DPA has proven to be a powerful
tool for the computation of radiative corrections. In order to
analyse, for the process considered here, the applicability of this
approximation in a wide range of energies, we have plotted in
\reffi{fi:WZ_born_pt} the tree-level cross section as a function of
the $\PT(l'\bar l')$ cut.
The first three curves represent, from top to bottom, the contribution
of the pure doubly-resonant (DR) diagrams, the full result including
all Feynman diagrams which contribute to the same final state (the
number of diagrams is 10 in absence of identical particles in the
final state, otherwise it doubles), and finally the DPA as defined in
\refeq{eq:BornWZ}.

If one looks at the difference between the complete result and the
DPA, one can see that the discrepancy is rather remarkable.  It
amounts in fact to roughly 15$\%$ for $\PT(l'\bar l')$ cuts above
$100\GeV$.  Let us note that, for this process, the commonly adopted
narrow-width approximation, which corresponds to {\it{production
    $\times$ decay}}, differs from the DPA by less than 2$\%$ for the
considered range of $\PTcut(l'\bar l')$.  So, depending on applied cuts
and selected energy range, the narrow-width approximation can
underestimate the exact result by roughly 13$\%$.  This points out the
need of using the exact matrix element at lowest order, as specified
in \refeq{M2ddef}.

The second information one can extract from \reffi{fi:WZ_born_pt} is
related to the contribution of non-DR diagrams.  As shown by the
dashed-line, the DR contribution ($\Pp\Pp\to \PW\PZ\to 4f$), which is
lower than the exact result by about 1$\%$ around threshold, increases
with energy relatively to the full result.  For
$\PTcut(l'\bar l')=300\GeV$, the difference between the two cross
sections is already of order 20$\%$, and at very large energies the DR
diagrams can even overestimate the result by a factor 2 or more.  This
effect is due to delicate gauge cancellations between DR and non-DR
diagrams, which characterize the behaviour of off-shell cross sections
in the high-energy regime.  DR and non-DR diagrams do not constitute
two separately gauge-independent subsets. Hence, the pure DR
contribution cannot be considered as a physical observable and the
signal definition based on the diagrammatic approach and commonly
adopted for example at LEP2 for $\PW\PW$ and $\PZ\PZ$ physics is not
anymore adequate to describe di-boson production at the LHC in the
high-$\PT$ region. The only sensible observable is the total
contribution or the DPA which is a well-defined gauge-independent
quantity.
 
\begin{figure}
  \begin{center}
\epsfig{file=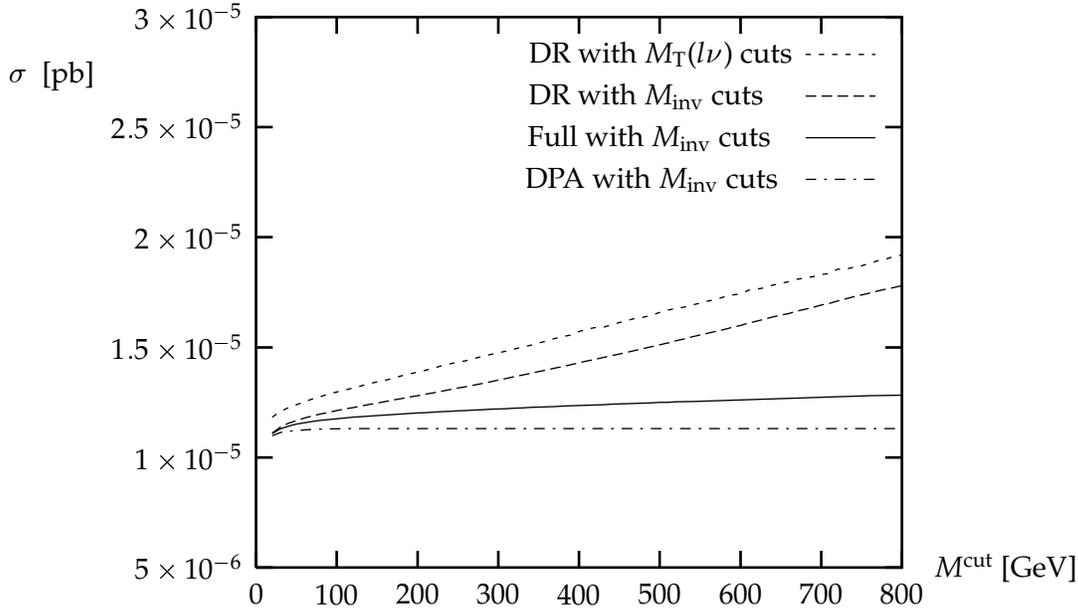, angle=0}
\end{center}  
\caption{Born cross section for the process $\Pp\Pp\to
  l\nu_ll^\prime\bar{l^\prime}$ at $\sqrt{s}=14\TeV$ as a function of
  the upper cut $\Mcut$ on the two invariant masses, $M(ij)$, of the
  leptonic pairs which could reconstruct $\PW$ and $\PZ$ bosons, as
  explained in the text.  Standard cuts and
  $\PT(l^\prime\bar{l^\prime})>800\GeV$ are applied.}
\label{fi:WZ_born_mcut}
\end{figure}

In order to investigate whether the difference between DR and full
result is essentially due to the off-shellness of the gauge bosons as
expected, we have then studied the effects of possible kinematical
cuts. In \reffi{fi:WZ_born_mcut}, we have plotted the cross section
for the extreme case $\PT(l'\bar l')=800\GeV$ as a function of an upper
cut $\Mcut$ applied on the two invariant masses $M(ij)$ of the
leptonic pairs which could reconstruct the $\PZ$ and $\PW$ bosons,
$M(ij) < M_V+\Mcut$. We assume a lower cut $M(ij)>M_V-20\GeV$, which
is kept fixed in order to suppress the contribution from the virtual
photon.
As can be seen, for $\Mcut=20\GeV$ the difference between DR and the
exact result reduces to the per-cent level. Also, both converge
towards the DPA value, represented by the nearly flat dot-dashed line.
It is quite obvious that cross sections computed in DPA are not
sensible to this kind of cuts, as the gauge bosons are always
considered on-shell except for the weakly cut-dependent factor $\left
  [(p_\PW^2-\MW^2)^2+\GW^2\MW^2\right ]^{-1}\left
  [(p_\PZ^2-\MZ^2)^2+\GZ^2\MZ^2\right ]^{-1}$, which reproduces the
resonant peaking structure.

Of course, a cut on the invariant mass of the $l\nu$ pair is not
physical since the longitudinal momentum of the neutrino is not
directly measurable.  We have therefore imposed the same kind of cuts,
but using the transverse mass $\MT(l\nu )=\sqrt{\ET^2(l\nu
  )-\PT^2(l\nu )}$ as physical quantity instead of the $M(l\nu )$
invariant mass and releasing the lower cut on $\MT(l\nu )$.  The
conclusion is similar. The DR contribution differs at the order of ten
per cent from the previous case, as shown by the dotted curve in
\reffi{fi:WZ_born_mcut}. The full calculation, which represents the
true observable, and the DPA are instead rather insensitive to this
change.
 
In the following, we assume the additional kinematical cuts
\beq\label{addcuts}
\MT(l\nu )< \MW +20\GeV,\qquad 
|M(l^\prime\bar{l^\prime} )-\MZ|< 20\GeV
\end{equation}
under which exact result and DPA coincide at per-cent level, with a
modest loss of signal, as shown in \reffi{fi:WZ_born_pt} where the
lower solid line represents the full result after imposing the
above-mentioned cuts. Since the exact cross section for the process
$\Pp\Pp\to 4f$ is rather well approximated by the DPA if proper cuts
are applied, we can safely adopt the DPA for computing electroweak
radiative corrections.  Let us notice, however, that electroweak
radiative contributions are not much larger than 20$\%$ in the region
of experimental sensitivity, as shown in the next section, and the DPA
differs from the exact result by less than 15$\%$. Therefore, without
imposing the additional cuts \refeq{addcuts}, the error induced by use
of the DPA in computing $\O(\alpha )$ contributions would give rise to
an uncertainty of less than 3$\%$ on the total cross section, so well
below the statistical accuracy.

\subsection{\boldmath Effects of $\O(\alpha )$ corrections}
\label{se:WZRC}

In this subsection, we discuss the effect of leading-logarithmic
electroweak virtual corrections to $\PW\PZ$ production in DPA. First
of all, one can see in \reffi{fi:WZ_born_inv}, where the distributions
in the reconstructed invariants $\sqrt{\hat{s}}$ and $\sqrt{|\hat r|}$
are plotted, that the previously discussed conditions, under which the
logarithmic high-energy approximation is valid, are well fulfilled for
$\PW\PZ$ production at high transverse momentum $\PT(l\bar l)$. Both
the hard-scattering invariant mass $\sqrt{\hat s}$ and
$\sqrt{|\hat{r}|}=\sqrt{|x_iP_i-p(l\bar l)|^2}$, where $x_iP_i$ is the
momentum of the parton from one of the protons ($\hat{r}$ corresponds
to $\hat{t}$ or $\hat{u}$ depending on the partonic process), are in
fact much larger than the boson masses.
We have checked in addition that most part of the contribution to the
cross section comes from the region where the scattering angle of the
reconstructed $\PZ$~boson in the $\PW\PZ$ rest-frame is in the central range 
with respect to the beam.  
Finally, as to the ratios between the different invariants which appear in 
the logarithms, we have verified that the pure angular-dependent ones are 
mostly in the range $1<\hat s/|\hat{r}|<6$, while $\hat s/\MW^2>50$ thus 
allowing to omit $\log^2{(|\hat{r}|/{\hat s})}$ type of logarithms up to an 
accuracy of a few per cent.

As already mentioned in \refse{se:ewrc}, we perform the computation of
radiative corrections to the full process $\Pp\Pp\to 4f$ in DPA, using
the complete expressions given in \refapp{app:corr}, i.e. implementing
the full (SU(2)-transformed) Born matrix elements as in
\refeq{SCWZcorr}, \refeq{duSCgen}, \refeq{SSCcorrections}, and
\refeq{CcorrectionsL}--\refeq{eq:deCPRTT}.  We have verified that the
results obtained by making use of the high-energy approximation for
the Born amplitudes given in \refeq{longbornWN} and \refeq{trabornWN}
are in very good agreement.  For all results given in the following,
the difference between the two methods is in fact at per-mille level.
This comparison shows the reliability of the high-energy approximation
for the Born matrix elements, under which the correction factor can be
factorized and expressed in a very compact and simple form, leading to
considerable decrease of CPU time.

\begin{figure}
\begin{center}
\epsfig{file=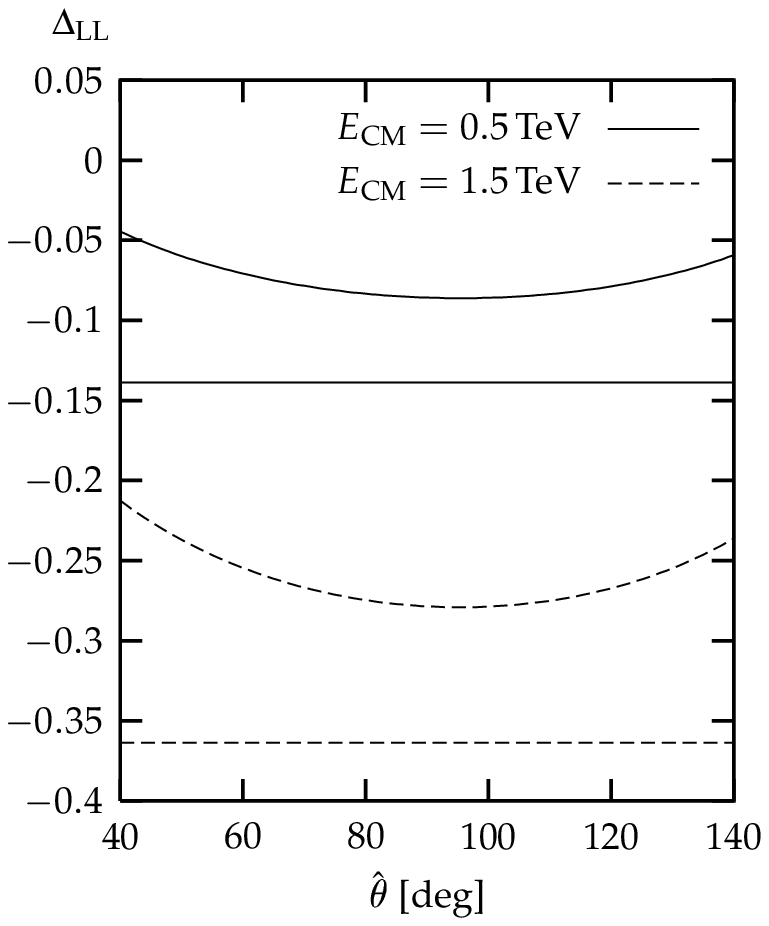, angle=0}
\epsfig{file=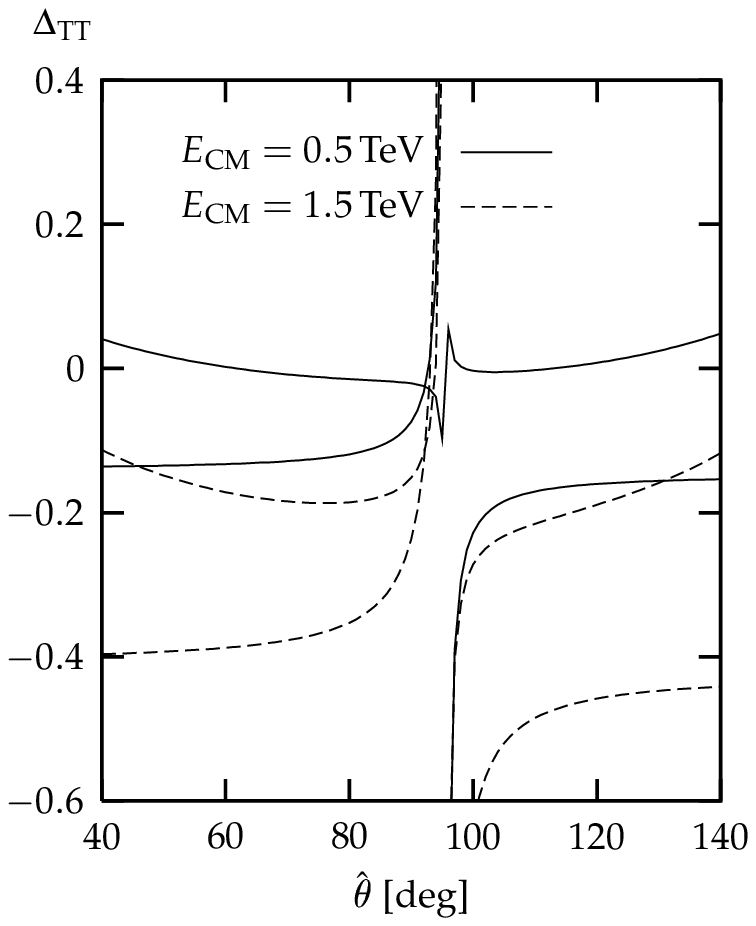, angle=0}
\end{center}
\caption{(a) Relative corrections $\Delta_{\LL}$ to the
  angular distributions for the process
  $\bar\Pd\Pu\rightarrow\PW^+_{\rL}\PZ_{\rL}$.  For each CM energy,
  the upper curves include the complete logarithmic corrections, the
  lower curves only the angular-independent logarithms.  (b) The same
  for the process $\bar\Pd\Pu\rightarrow\PW^+_{\rT}\PZ_{\rT}$.}
\label{fi:WZ_corr_theta}
\end{figure}

In order to discuss the basic structure of radiative corrections, we
first consider the $\O(\alpha )$ contributions to the partonic
subprocess $\bar\Pd\Pu\to\PWp\PZ$. In \reffi{fi:WZ_corr_theta}a we
plot the relative correction to the angular distribution of the
longitudinal component
$\Delta_{\LL}=(\rd\sigma^{\LL}/\rd\cos\hat\theta
-\rd\sigma_{\Born}^{\LL}/\rd\cos\hat\theta)/
(\rd\sigma_{\Born}^{\LL}/\rd\cos\hat\theta )$ with $\sigma =\sigma
(\bar\Pd\Pu\to\PW^+\PZ)$ as a function of the angle $\hat\theta$
between the $\PZ$~boson and the $\bar\Pd$~quark in the $\bar\Pd\Pu$ CM
frame for the two energies $E_{\CM}=0.5\TeV$ and $1.5\TeV$. As can be
seen, the $\LL$ part receives sizeable corrections, in particular in
the central region $\hat\theta\simeq 90^\circ$ where $\sigma_{\LL}$ is
more enhanced.  In order to pinpoint the effect of the
angular-dependent contributions to the radiative corrections, in the
same figure we have also plotted the two flat curves which include
only angular-independent logarithms of $\hat{s}/\MW^2$. The difference
between the two results for each CM energy shows the importance of
taking into account leading and full subleading terms. There are in
fact partial cancellations occurring between angular-dependent and
angular-independent parts, which sizeably lower the overall
corrections (see also \citere{Beccaria:2001yf}).

For the transverse part $\sigma_{\TT}$ the corresponding relative
corrections $\De_{\TT}$ are shown in \reffi{fi:WZ_corr_theta}b for two
values of $E_{\CM}$ as in the previous figure. Here,
radiative-correction effects are less pronounced compared to the $\LL$
case, especially at extreme angles where $\sigma_{\TT}$ receives its
maximal contribution. The spikes in \reffi{fi:WZ_corr_theta}b
originate from the radiation zero of the lowest-order cross section
(see \reffi{fi:WZ_born_theta}), the absolute corrections behave
smoothly everywhere. The angular behaviour of $\De_{\TT}$ is more
complex, compared with the longitudinal one. The dependence on the
angle has in fact a two-fold origin.
In addition to the angular-dependent logarithms [see
\refeq{SSCcorrections} and \refeq{SSCcorr2}], there are
angular-independent double logarithms $\log^2(\hat{s}/\MW^2)$ with
angular-dependent coefficients [see \refeq{duSCgen} and
\refeq{duSCgen2}], which originate from the mixing of the final
\PZ~boson with the photon, induced by virtual soft--collinear
\PW~bosons.  In \reffi{fi:WZ_corr_theta}b we have reported the total
deviation $\Delta_{\TT}$, represented for each CM energy by the upper
curves, and the partial contribution coming from the
angular-independent logarithms of $\hat{s}/M^2_W$, given by the lower
lines. As can be seen, also in this case, the correction factors
proportional to the angular-dependent and angular-independent
logarithms have opposite sign, leading as before to a reduction of the
total correction.

\begin{figure}
  \begin{center}
\epsfig{file=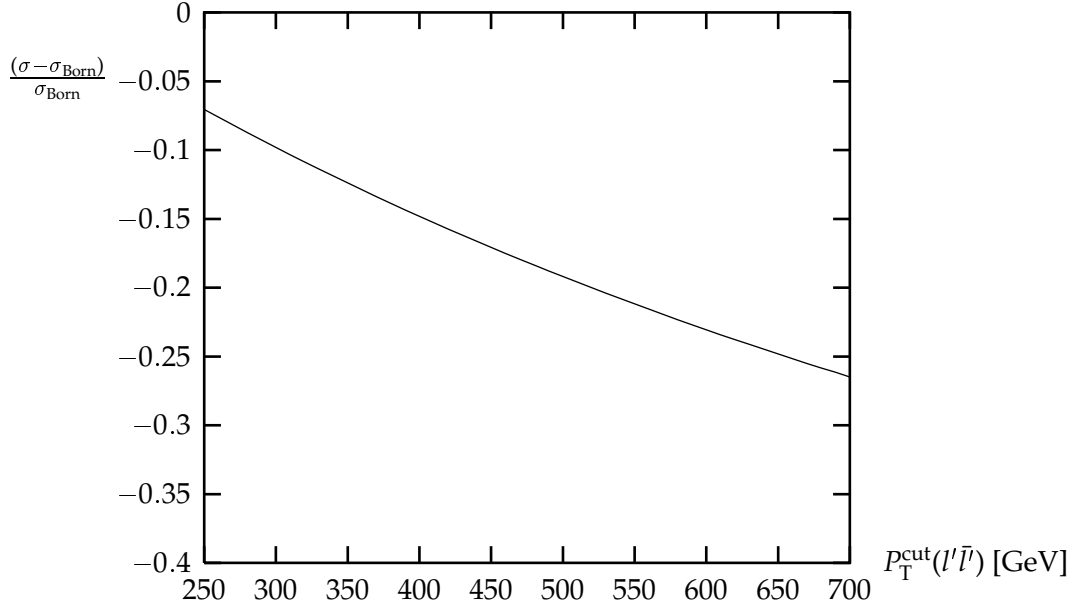, angle=0}
\end{center}  
\caption{Relative correction to the total cross section for the full process
  $\Pp\Pp\to l\nu_ll^\prime\bar{l^\prime}$ at $\sqrt{s}=14\TeV$ as a
  function of the cut on the reconstructed $\PZ$-boson transverse
  momentum. Standard cuts are applied.}
\label{fi:WZ_corr_pt}
\end{figure}

In order to show the effect of the electroweak radiative corrections
on the complete process $\Pp\Pp\to 4f$, in \reffi{fi:WZ_corr_pt} we
have plotted the $\O(\alpha )$ correction relative to the total Born
cross section, $\Delta =(\sigma -\sigma_{\Born})/\sigma_{\Born}$, as a
function of the cut on the transverse momentum of the reconstructed
$\PZ$~boson, $\PTcut(l\bar l)$.
Our standard cuts are applied. As can be seen, the $\O(\alpha )$
contributions are negative and get larger with increasing $\PTcut$,
roughly going from $-5\%$ to $-25\%$ in the considered momentum range.
The relatively large size of the radiative corrections, especially at
energies which are at first sight rather modest (\eg $\PTcut>250\GeV$
implies $\sqrt{\hat s}>500\GeV$), is mainly due to two combined
effects. On one side, the longitudinal component of the cross section,
$\sigma_{\LL}$, which dominates at low values of the allowed energy
range, as shown in \reffi{fi:WZ_born_ecm} (right side), where the
scattering angles are dominantly central, generates sizeable
corrections. On the other hand, as can be seen in
\reffi{fi:WZ_born_inv}, the total cross section even for modest values
of $\PTcut$ receives a substantial contribution from the
very-high-energy region, where $\sigma_{\TT}$ dominates.  So, the
generally smaller $\O(\alpha )$ contributions from the $\TT$
configuration, as compared with the $\LL$ ones, get enhanced by the
higher values of the CM energy, and give globally additional sizeable
effects.  As a consequence, the corrections to the total cross section
are large because the $\PT$ cut selects high-energy domains (since the
only way to obtain a large $\PT(l\bar l)$ is to have a large $\PW\PZ$
invariant mass) and enhances the contributions coming from the central
angular region.  This shows that the size of the radiative
contributions is strictly dependent on the applied cuts and the
selected kinematical configurations.

\begin{figure}
  \begin{center}
\epsfig{file=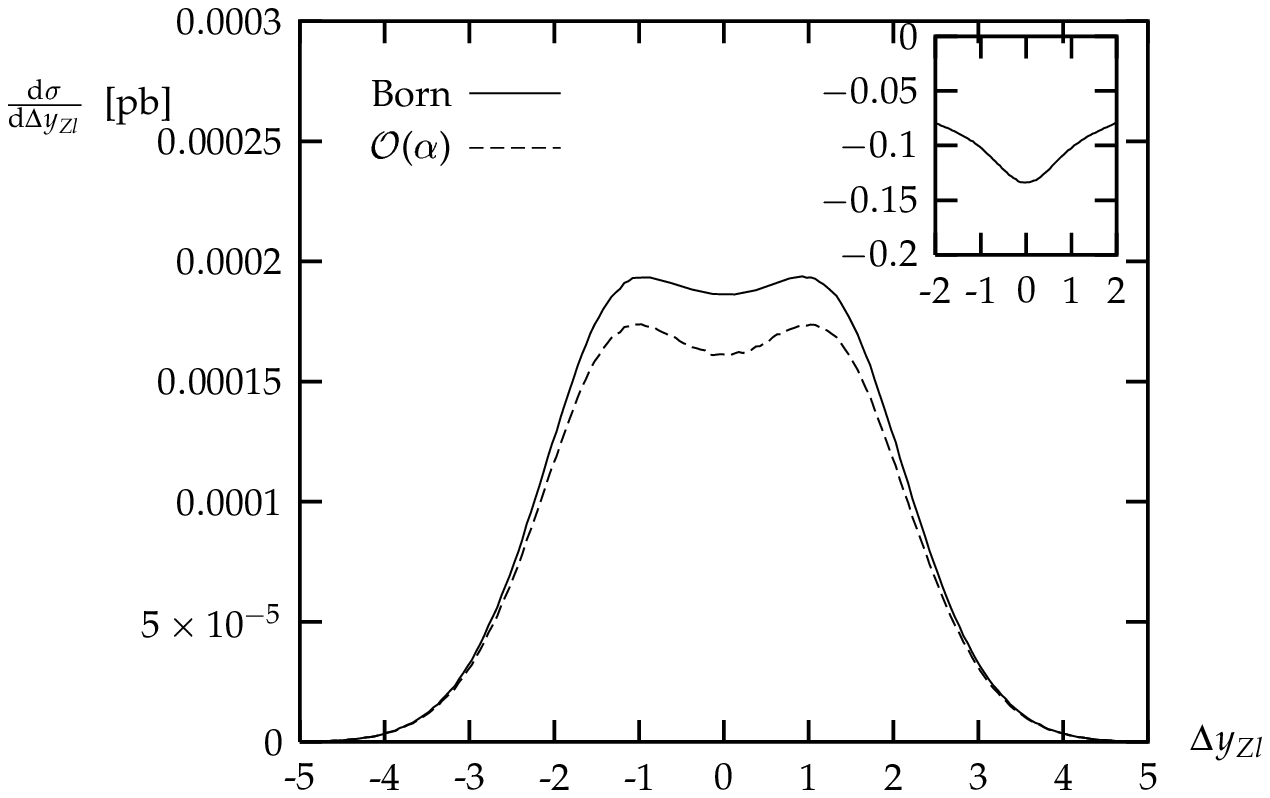, angle=0}
\end{center}  
\caption{Rapidity distribution for the full process
        $\Pp\Pp\to l\nu_ll^\prime\bar{l^\prime}$ at $\sqrt{s}=14\TeV$.
        Standard cuts and $\PT(l^\prime\bar{l^\prime})>300\GeV$ are
        applied. The inset plot shows the difference between $\O(\alpha
        )$ and Born results normalized to the Born distribution.}
\label{fi:WZ_corr_ylZ}
\end{figure}

This is even more clearly visible in \reffi{fi:WZ_corr_ylZ}, where we
have plotted the distribution in the difference between the rapidity
of the reconstructed $\PZ$~boson and of the charged lepton coming from
the decay of the $\PW$, $\Delta y_{\PZ l}=y(l^\prime\bar{l^\prime})-y(l)$,
at Born level (solid line) and including radiative corrections (dashed
line).
The rapidity is defined from the energy $E$ and the longitudinal
momentum $P_{\rL}$ by $y=0.5\log ((E+P_\rL)/(E-P_\rL))$.  This
variable, studied in \citeres{Baur:1995aj,Dixon:1999di} and defined in terms of
direct observables, is symmetric around zero and shows a residual dip
reflecting the approximate radiation zero of the angular distribution
of the Born $\PW\PZ$ production. The quantity $\Delta y_{\PZ l}$ is in
fact similar to the rapidity difference $\Delta y_{\PZ\PW}=y_\PW
-y_\PZ$ considered in \citere{Frixione:1992pj}, which is strictly
related to the scattering angle, $\hat\theta$, of the $\PZ$~boson in
the $\PW\PZ$ rest-frame. The definition of $\Delta y_{\PZ\PW}$ and
$\cos\hat\theta$ requires, however, the reconstruction of the unknown
longitudinal momentum of the neutrino.  Even if this can be derived by
assuming the $\PW$~boson to be on-shell \cite{nurec}, the two-fold
ambiguity given by the two possible solutions for the neutrino
longitudinal momentum spoils the radiation zero. Therefore, in order
to extract informations about the angular dependence of the $\PW\PZ$
process, it is preferable to use $\Delta y_{\PZ l}$
\cite{Baur:1995aj,Dixon:1999di}.

The first information one can get from \reffi{fi:WZ_corr_ylZ} is that
the main contribution to the cross section originates from small
values of $\Delta y_{\PZ l}$ corresponding to central scattering
angles.  At the LHC, for the first time the statistics will be
sufficient to experimentally test the behaviour due to the approximate
radiation zero, which might be distorted by new-physics contributions.
Figure \ref{fi:WZ_corr_ylZ} indicates that radiative effects are
maximal at small rapidity separation, which is the region of stronger
sensitivity to new physics. Moreover, owing to the applied cuts, these
relatively large radiative contributions are not due to the
suppression of the tree-level cross section and, being negative, they
even slightly enhance the residual dip.

The above-discussed effects should of course be compared with the
expected experimental accuracy. In \refta{ta:WZ} we have listed the
relative deviation $\Delta$ and the statistical error, estimated by
assuming a luminosity $L=100 \fba^{-1}$ for two experiments, for some
$\PTcut$ values.  This comparison indicates that at high transverse
momentum of the gauge bosons the virtual electroweak corrections are
non-negligible and can be comparable with the experimental accuracy up to
about $500 \GeV$. In this region the corrections range between $-5$ and 
$-20\%$.
\begin{table}\centering
$$
\begin{array}{|c|c|c|c|c|}
\hline 
\multicolumn{5}{|c|}{\Pp\Pp\to
  l\nu_ll^\prime\bar{l^\prime}\rule[-2ex]{0ex}{5ex}}\\
\hline
\PTcut(l^\prime\bar{l^\prime})~[\mathrm{GeV}] & \sigma_{\Born}~[\mathrm{fb}] & 
~~\sigma ~[\mathrm{fb}]~~ &~\Delta~[\%]~~ & 1/\sqrt{2L\sigma_{\Born}}~[\%] \\
\hline
\hline
250 & 1.716 & 1.595 & ~{-7.1} & 5.4 \\
\hline
300 & 0.899 & 0.811  & -9.8  & 7.5 \\
\hline
350 & 0.503 & 0.441 & -12.4 & \phantom{.}10 \\
\hline
400 & 0.296 & 0.252 & -14.9 & \phantom{.}13 \\
\hline
450 & 0.181 & 0.150 & -17.1 & 16.6 \\
\hline
500 & 0.114 & 0.092 & -19.3 & 20.9 \\
\hline
\end{array}$$
\caption {Cross section for $\Pp\Pp\to l\nu_l l^\prime \bar{l^\prime}$ 
for various values of $\PTcut(l^\prime\bar{l^\prime})$} 
\label{ta:WZ}
\end{table}
Whether or not they should be taken into account when performing
analyses in this kinematical region depends of course on the available
luminosity.  Only in a high-luminosity run their effect will be
relevant.

\section {\boldmath $\PW^\pm\gamma$ production}
\label{se:WA}

In this section, we extend our analysis to the process $\Pp\Pp\to
l\nu_l\gamma$ ($l=\Pe,\mu$). This channel, proper for the measurement
of the trilinear gauge-boson coupling $\PW\PW\gamma$, can furnish
complementary informations on the vertex structure of the SM when
combined with the analysis of $\PW\PZ$ production.  As before, we
consider the region of high CM energies of the hard scattering, where
the sensitivity to new-physics effects is expected to be more
enhanced, and the precise knowledge of the SM background can be then
particularly useful. In the following we analyse the same set of
variables used to discuss the $\PW\PZ$ production process and the
effect of the $\O(\alpha )$ electroweak corrections on them.

\subsection{Born level}
\label{se:WA:Born}

The partonic subprocess $\bar\Pd\Pu\rightarrow \PWp\gamma$ is
dominated by the production of two transverse gauge bosons, whereas
the remaining ($\LT$) helicity configuration is suppressed by a factor
$\MW/E_\CM$ in the high-energy limit.  All features discussed in
\refse{se:WZ:Born} for the subprocess $\bar\Pd\Pu\rightarrow
\PW^+_{\rT}\PZ_{\rT}$ with transversally polarized gauge bosons
qualitatively apply as well to $\PW\gamma$ production. The
corresponding cross section is in fact strongly peaked in the forward
and backward directions and presents a radiation zero for
$\cos\hat\theta =(Q_\Pu+Q_\Pd)/(Q_\Pu-Q_\Pd)=1/3$ where $\hat\theta$
is the angle between the $\bar\Pd$~quark and the photon. As to the
general behaviour of the $\bar\Pd\Pu\rightarrow\PW^+\gamma$ process,
we refer back to \reffis{fi:WZ_born_theta} and \ref{fi:WZ_born_ecm}
and details given in the text.

\begin{figure}
\begin{center}
\epsfig{file=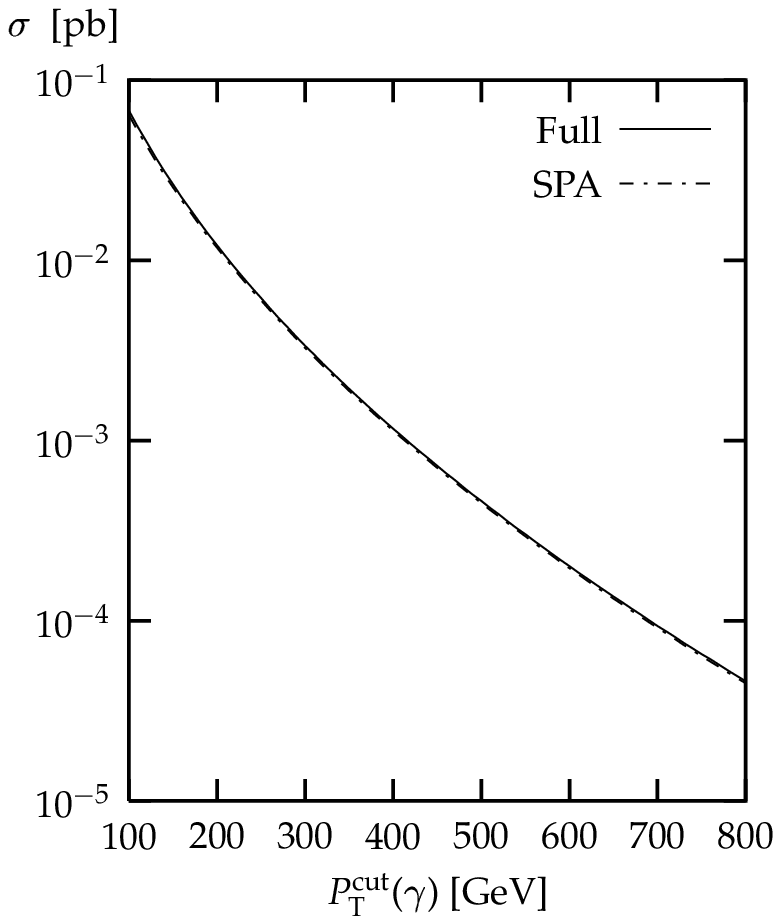, angle=0}
\epsfig{file=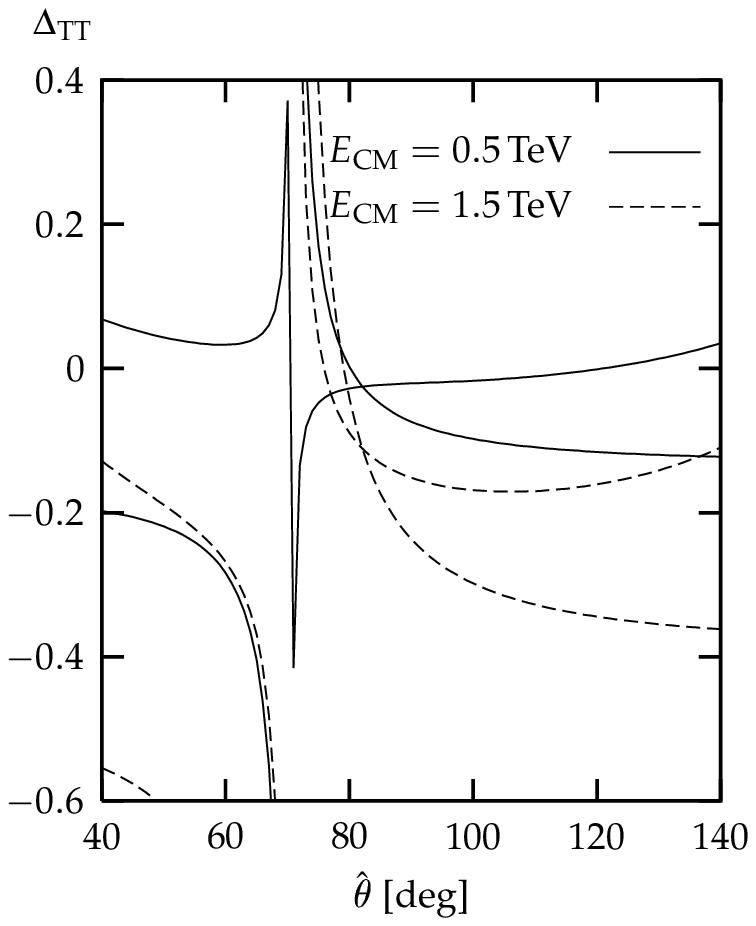, angle=0}
\end{center}
\caption{(a) Born cross section for the full process 
  $\Pp\Pp\to l\nu_l\gamma$ at $\sqrt{s}=14\TeV$ as a function of the
  cut on the photon transverse momentum. Standard cuts are applied.
  (b) Relative corrections to the angular distribution for the
  subprocess $\bar\Pd\Pu\rightarrow\PW^+_{\rT} \ga$. The two upper and
  lower curves for each energy are as in \reffi{fi:WZ_corr_theta}b.}
\label{fi:WA_corr_theta}
\end{figure}

In spite of these similarities, $\PW\gamma$ production presents,
however, some different characteristics with respect to the $\PW\PZ$
case. First of all, owing to the absence of any pure non-suppressed
longitudinal components, there are no sizeable gauge cancellations in
the total cross section of the full process $\Pp\Pp\rightarrow
l\nu_l\gamma$ at high energy. The resonant contribution ($\Pp\Pp\to
\PW\gamma\to l\nu_l\gamma$) is always lower than the full result, also
for high values of the cut on the photon transverse momentum
$\PT(\gamma )$, and the difference between the two cross sections is
below 3$\%$.  Therefore, one can still consider the pure resonant part
as a useful definition of the $\PW\gamma$ signal. Also, the
single-pole approximation (SPA) defined in \refeq{eq:BornWA} differs
negligibly from the exact result, as shown in
\reffi{fi:WA_corr_theta}a, where we have plotted the total cross
section versus the cut applied on the photon transverse momentum.
One could directly use the SPA to compute radiative corrections at the
per-cent level without imposing any additional cuts.  However, for
sake of uniformity, in the following we apply the same kind of cut
$\MT(l\nu )< \MW +20\GeV$ as used for the $\PW\PZ$ process, under
which resonant, full, and SPA cross sections converge to the same
value within one per cent and with a negligible loss of signal.

\subsection{\boldmath Effect of $\O(\alpha )$ corrections}

In this subsection, we study the effect of virtual electroweak
radiative corrections on $\PW\gamma$ production in SPA.  We consider
first the partonic subprocess $\bar\Pd\Pu\rightarrow\PWp\gamma$.  In
\reffi{fi:WA_corr_theta}b we have plotted
$\Delta_{\TT}=(\rd\sigma^{\TT}/\rd\cos\hat\theta -
\rd\sigma^{\TT}_{\Born}/\rd\cos\hat\theta ) /(
\rd\sigma^{\TT}_{\Born}/\rd\cos\hat\theta )$ as a function of the
angle $\hat\theta$ between the $\bar\Pd$~quark and the photon.  As can
be seen, the behaviour of the $\O(\alpha )$ contributions is quite
similar to the $\PW_{\rT}\PZ_{\rT}$ case. Only the spikes, again due
to the radiation zero, are correspondingly shifted, and reverse the
shape of the curves with respect to the angle, compared with
\reffi{fi:WZ_corr_theta}b.

\begin{figure}
  \begin{center}
\epsfig{file=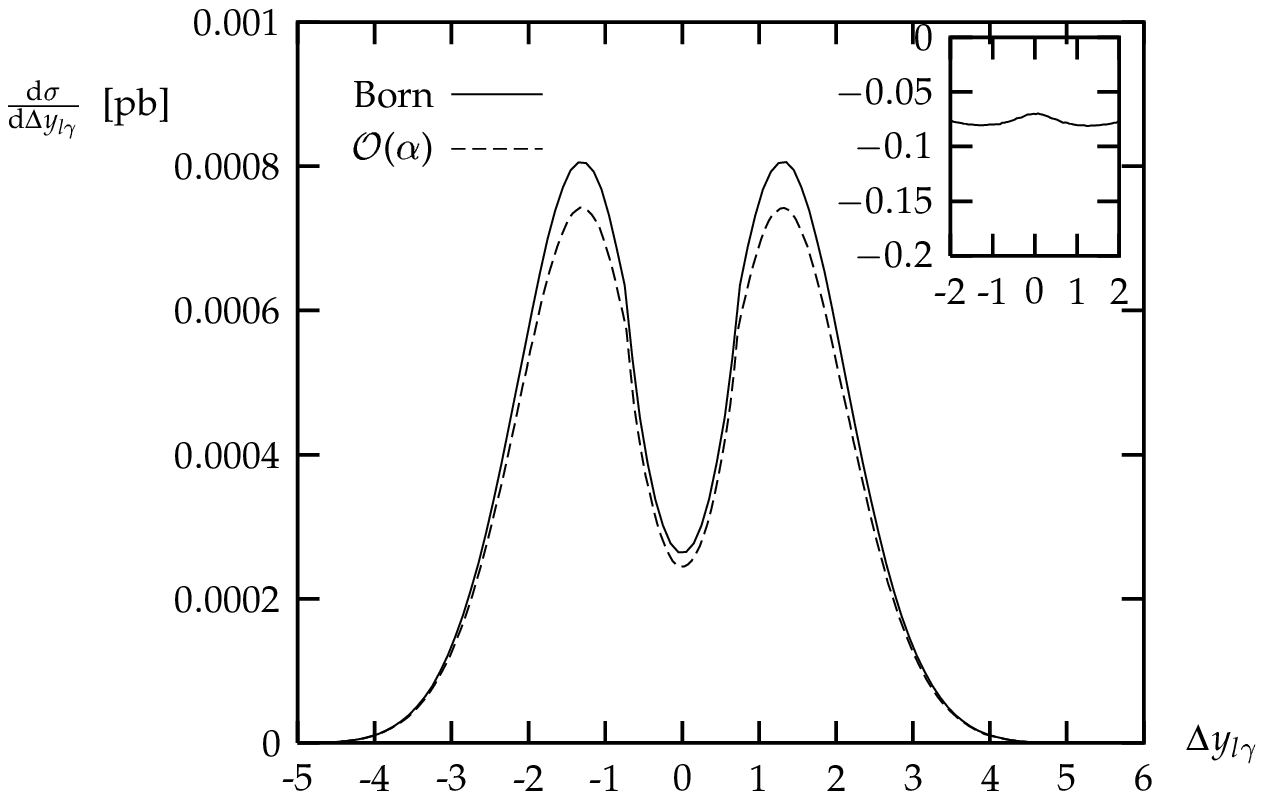, angle=0}
\end{center}  
\caption{ Rapidity distribution for the full process 
  $\Pp\Pp\to l\nu_l\ga$ at $\sqrt{s}=14\TeV$. Standard cuts and
  $\PT(\ga)>300\GeV$ are applied.  The inset plot shows the difference
  between $\O(\alpha )$ and Born results normalized to the Born
  distribution.}
\label{fi:WA_corr_ylZ}
\end{figure}

As in the previous case, the validity conditions of the high-energy
logarithmic approximation for the radiative corrections are well
satisfied by the complete process $\Pp\Pp\rightarrow l\nu_l\gamma$.
The kinematical behaviour of $\PW\gamma$ at high transverse momentum
of the photon reproduces in fact the same shape of the distributions
as in \reffi{fi:WZ_born_inv}. All invariants are then much larger than
the boson masses, and at fixed $\PTcut (\ga )$ the process receives
considerable contributions from very high CM energies. We have
moreover verified that, despite the radiation zero and the absence of
any non-mass-suppressed longitudinal components, for large $\PTcut
(\ga )$ values ($\PTcut (\ga )\gsim 250\GeV$) most part of the
contribution to the total cross section comes from the region of phase
space where the photon is emitted at large angle with respect to the
beam (see also \reffi{fi:WA_corr_ylZ}).

\begin{figure}
  \begin{center}
\epsfig{file=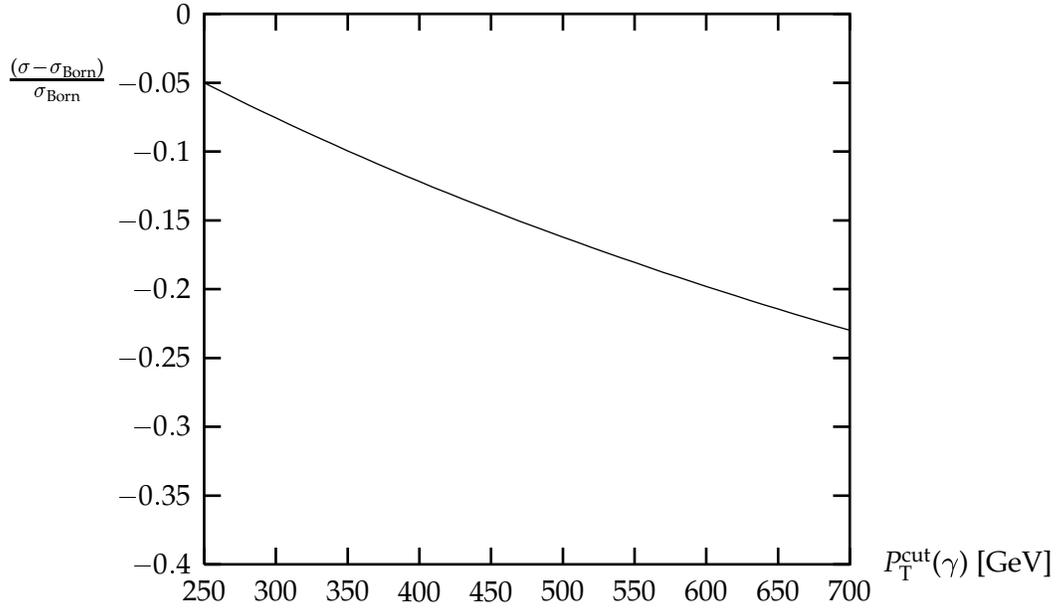, angle=0}
\end{center}  
\caption{Relative  corrections to the cross section for the full
  process $\Pp\Pp\to l\nu_l\ga$ at $\sqrt{s}=14\TeV$ as a function of
  the cut on the photon transverse momentum. Standard cuts are
  applied.}
\label{fi:WA_corr_pt}
\end{figure}

In order to show the effect of radiative corrections on the full
process $\Pp\Pp\rightarrow l\nu_l\gamma$, we have plotted as before
the $\O(\alpha )$ corrections relative to the Born cross section as a
function of the cut on the transverse momentum of the photon in
\reffi{fi:WA_corr_pt}.
The overall behaviour is quite similar to the $\PW\PZ$ case; the size
of the radiative effects is, however, lower.

This affects in the same way also the distribution in the difference
between the rapidity of the photon and the charged lepton coming from
$\PW$-boson decay \cite{Baur:1994sa}, $\Delta
y_{l\gamma}=y_l-y_\gamma$, plotted in \reffi{fi:WA_corr_ylZ}.
Here, unlike in the previous case, the dip reflecting the radiation
zero is much more pronounced, but the radiative corrections slightly
decrease going towards $\Delta y_{l\gamma}=0$. Owing to the different
location of the radiation zero, the radiative contribution for very
small rapidity separation is still sizeable, but it does not get
enhanced as for the $\PW\PZ$ process, where in the same region the
longitudinal component gives the dominant contribution. Hence for
$\PW\ga$, despite the complex behaviour shown in
\reffi{fi:WA_corr_theta}b, which is merely due to the fictitious
spikes, the radiative-correction effect is rather uniform in the
angular range we consider, and leads to an overall rescaling of the
$\Delta y_{l\gamma}$ distribution by roughly a factor 0.9.  These
effects could still mimic the behaviour of new-physics contributions.
Their smaller size, compared with the $\PW\PZ$ case, is compensated by
the larger value of the overall cross section.  Therefore, even if not
extremely enhanced in the central rapidity range, radiative effects
can become comparable with the statistical error.

In Tab.~2 we compare the \Oa~relative correction $\Delta$ to the Born
cross section with the expected experimental accuracy, assuming $L=100
\fba^{-1}$ for two experiments, for different values of $\PTcut (\ga
)$.
\begin{table}[hbt]\centering$$
\begin{array}{|c|c|c|c|c|}
\hline 
\multicolumn{5}{|c|}{\Pp\Pp\to l\nu_l\ga\rule[-2ex]{0ex}{5ex}}\\
\hline
\PTcut(\ga)~[\mathrm{GeV}] & \sigma_{\Born}~[\mathrm{fb}] & 
~\sigma ~[\mathrm{fb}]~ &~\Delta~[\%]~ & 1/\sqrt{2L\sigma_{\Born}}~[\%] \\
\hline
\hline
250 & ~5.810 & ~5.519 & ~{-5.0} & ~2.9 \\
\hline
300 & ~3.180 & ~2.940 & ~{-7.6} & ~4.0 \\
\hline
350 & ~1.832 & ~1.650 & -10.0 & ~5.2 \\
\hline
400 & ~1.100 & ~0.966 & -12.2 & ~6.7 \\
\hline
450 & ~0.684 & ~0.587 & -14.2 & ~8.6 \\
\hline
500 & ~0.437 & ~0.366 & -16.2 & 10.7 \\
\hline
550 & ~0.285 & ~0.234 & -18.0 & 13.2 \\
\hline
600 & ~0.190 & ~0.152 & -19.8 & 16.2 \\
\hline
650 & ~0.129 & ~0.101 & -21.4 & 19.6 \\ 
\hline
700 & ~0.089 & ~0.068 & -23.3 & 23.7 \\
\hline

\end {array}$$
\caption{Cross section for $\Pp\Pp\to l\nu_l\ga$ for various values of
  $\PTcut(\ga)$.} 
\label{ta:WA}
\end {table}
As one can see, radiative effects are very sensitive to $\PTcut (\ga
)$ and, despite of the decrease of the cross section with increasing
$\PTcut (\ga )$, are larger than the statistical error for $\PTcut(\ga)$ 
below $700 \GeV$, where they range from $-5$ to $-23\%$. Moreover, they
could be of some relevance also in a low-luminosity run ($L=30
\fba^{-1}$) of the LHC, as they might become comparable with the experimental
precision for $\PTcut (\ga )< 400 \GeV$.
\section{Conclusion}
\label{se:concl}

By means of a complete four-fermion calculation, we have examined
$\PW\PZ$ production in the purely leptonic channel at the LHC. An
analogous computation has been performed for the $\PW\ga$ process
followed by the leptonic $\PW$ decay. We have given some examples of
phenomenological analyses relevant to hadronic di-boson production in
the high di-boson invariant-mass region.

At tree level, we have found that, for processes involving $\PW\PZ$
production, the diagrammatic approach usually adopted to isolate the
signal is not viable anymore at large transverse momentum of the
reconstructed $\PZ$ boson, owing to gauge cancellations. The
doubly-resonant approximation can differ from the full result by tens
of per cent in experimentally relevant regions.  The only sensible
observable is the total contribution.  Moreover the two commonly used
approximations, narrow-width (i.e. {\it production $\times $decay})
and leading pole approximation, can underestimate the exact result by
about 10--15$\%$ at relatively modest energies, if no cuts are applied.
   
The primary aim of our analysis was to investigate the structure of
virtual electroweak corrections and their effect on di-boson
production processes at the LHC. The one-loop leading-logarithmic
corrections to the full four-fermion or two-fermion-plus-photon process
have been calculated in leading-pole approximation, neglecting
non-factorizable corrections, and restricting oneself to the
gauge-invariant leading-logarithmic corrections, which only contribute
to the gauge-boson pair-production subprocess. We found that this
approach constitutes a reliable approximation in the high-$\PT$ region
at the LHC. 

In order to illustrate the behaviour
and the size of $\O(\alpha )$ contributions, we have presented
different cross sections and distributions. In this study, we have not
included the full QED radiative contributions, which involve also the
emission of real photons and therefore depend on the detector
resolution. We focused instead on the contributions of the leading
electroweak logarithms resulting from above the electroweak scale.

For $\PW\PZ$ and $\PW\ga$ production processes, electroweak
corrections turn out to be non-negligible in the high-energy region of
the hard process, in particular for large transverse momentum and
small rapidity separation of the reconstructed vector bosons, which is
the kinematical range of maximal sensitivity to new-physics phenomena.
Electroweak radiative effects lower the Born results by 5--20\% in the
region of experimental sensitivity. We have moreover shown that their
size depends sensibly not only on the CM energy but also on the
applied cuts and varies according to the selected observables and
kinematical regions. Despite of the strong decrease of the cross
section with increasing di-boson invariant mass, radiative effects can
still be appreciable if compared with the expected experimental
precision. This depends of course on the available luminosity. For
$\PW\PZ$ production, these effects are relevant for the
high-luminosity run of the LHC. Owing to their larger overall cross
section, $\PW\ga$ production processes can instead show a sensitivity
to radiative effects also at low luminosity.

\section*{Acknowledgements}
This work was supported in part by the Swiss Bundesamt f\"ur Bildung
und Wissenschaft and by the European Union under contract
HPRN-CT-2000-00149.

\begin{appendix}

\section{Logarithmic electroweak corrections}
\label{app:corr}

In this section, we present the analytical formulas for the
logarithmic electroweak corrections to the polarized partonic
subprocesses
\newcommand{\duwz}{\bar{\Pd}_\rL\Pu_\rL \to \PW^+_{\la_\PW} \PZ_{\la_\PZ}}
\newcommand{\duwn}{\bar{\Pd}_\rL\Pu_\rL \to \PW^+_{\la_\PW} N_{\la_N}}
\beq\label{duprocess}
\bar{\Pd}_\rL(p_{\bar{\Pd}})\,
\Pu_\rL(p_\Pu)
 \to 
\PW^+_{\la_\PW}(p_\PW)\,
 N_{\la_N}(p_N),\qquad N=A,\PZ,
\eeq
which can be derived from the general results given in
\citere{Denner:2001jv}.  The label $\rL$ indicates the left-handed
chirality of the initial-state quarks (right-handed quarks are not
considered since they cannot produce \PW~bosons), and
$\la_{\PW,N}=0,\pm 1$ represent the gauge-boson helicities.  The
photon field is denoted by $A$ in this appendix. The Mandelstam
variables read $\hat{s}=(p_{\bar{\Pd}}+p_\Pu)^2$,
$\hat{t}=(p_{\bar{\Pd}}-p_\PWp)^2$, and
$\hat{u}=(p_{\bar{\Pd}}-p_N)^2$, where the momenta of the initial and
final states are incoming and outgoing, respectively. In the
high-energy limit, we have $\hat{t}\sim
-\hat{s}(1+\cos{\hat\theta})/2$ and $\hat{u}\sim
-\hat{s}(1-\cos{\hat\theta})/2$, where $\hat\theta$ is the angle
between $\vec{p}_{\bar{\Pd}}$ and $\vec{p}_N$, in the CM frame of the
scattering quarks.

For the calculation of the cross section \refeq{eq:convol} we need
besides the process $\bar\Pd\Pu\rightarrow\PW^+\PN$ also its charge
conjugate and the cross sections for exchanged initial quarks. These
latter can be obtained from one another just by exchanging the
invariants $\hat{t}\leftrightarrow\hat{u}$.  Owing to CP invariance,
the charge-conjugate processes can be instead obtained from the
initial processes by applying a parity transformation,
\beq\label{CPsymm}
\M\left [\Pd(p_{\Pd})\bar \Pu(p_{\bar \Pu})\rightarrow\PW^-(p_{\PW})\PN(p_{\PN})\right ]=
\M\left [\bar \Pd(\tilde{p}_{\bar \Pd})\Pu(\tilde{p}_{\bar \Pu})\rightarrow\PW^+(\tilde{p}_{\PW})\PN(\tilde{p}_{\PN})
\right ]
\eeq
with ${\tilde p}=(E,-{\vec p})$ for ${p}=(E,{\vec p})$. So, also in
this case, the correction factors can be obtained from the same
initial process \refeq{duprocess}. The formulas we give in the
following for the process \refeq{duprocess} can therefore cover all
contributions we need for the complete $\PW\PZ$ and $\PW\gamma$
production processes.
 
The one-loop corrections are evaluated in the limit
\beq\label{sudaklim}
\hat{s}\sim \hat{t} \sim \hat{u} \gg \MW^2,
\eeq
and we restrict ourselves to the combinations of gauge-boson
helicities that are not mass-suppressed compared with $\sqrt{\hat{s}}$ in
this limit.  These correspond to the purely transverse and opposite
final state $(\la_\PW,\la_N)=(\pm,\mp)$, which we denote by
$(\la_\PW,\la_N)=(\rT,\rT)$, and, in the case of $\PWpm\PZ$
production, also to the purely longitudinal final state
$(\la_\PW,\la_\PZ)=(0,0)$, which we denote by
$(\la_\PW,\la_\PZ)=(\rL,\rL)$.

\subsection{One-loop corrections}
\label{app:one-loopcorr}

In the following, we present the results as relative corrections
\beq \label{relco}
\de_{\bar{\Pd}_\rL\Pu_\rL \to \PW^+_{\la} N_{\la}}
=\frac{
\de \M_{\virt}^{\bar{\Pd}_\rL\Pu_\rL \to \PW^+_{\la} N_{\la}}
(p_{\bar{\Pd}},p_\Pu,p_\PW,p_N )}
{\M_{\Born}^{\bar{\Pd}_\rL\Pu_\rL \to \PW^+_{\la} N_{\la}}
(p_{\bar{\Pd}},p_\Pu,p_\PW,p_N)}
\eeq
to the Born matrix elements. A more detailed derivation can be found
in \citere{PhDthesis}.

As shown in \citere{Denner:2001jv}, in the high-energy logarithmic
approximation the longitudinal gauge bosons can be replaced by the
corresponding would-be Goldstone bosons. Therefore, in our results for
longitudinal final states ($\la=\rL$), the substitutions
$\PW^\pm_\rL\to \phi^\pm$ and $\PZ_\rL \to \chi$ have to be performed.

The corrections \refeq{relco} are split as
\beq
\de=\de^{\mathrm{LSC}}+\de^{\mathrm{SSC}}+\de^{\mathrm{C}}+\de^{\mathrm{PR}}
\eeq
into leading ($\de^{\mathrm{LSC}}$) and subleading
($\de^{\mathrm{SSC}}$) contributions originating from soft--collinear
gauge bosons, contributions $\de^{\mathrm{C}}$ that originate from
collinear (or soft) gauge bosons and from wave-function
renormalization, and contributions $\de^{\mathrm{PR}}$ that originate
from parameter renormalization.  All these corrections are evaluated
in logarithmic approximation, \ie including all terms that involve
logarithms of the form $\log({\hat{s}}/{\MW^2})$ in the high-energy
limit.  More precisely, we restrict ourselves to
\begin{itemize}
\item the angular-independent double- and single-logarithmic
  corrections of the type $\alpha \log^2{(\hat{s}/\MW^2)}$ and
  $\alpha\log{(\hat{s}/\MW^2)}$, which involve only the ratio of the
  CM energy to the \PW-boson mass,

\item the double logarithms of the form 
  $\alpha\log{(\hat{s}/\MW^2)}\log(\MZ^2/\MW^2)$, and 

\item the angular-dependent double logarithms of
  the type $\alpha\log{(\hat{s}/\MW^2)}\log{(|\hat{r}|/\hat{s})}$, with 
  $\hat{r}=\hat{t},\hat{u}$.
\end{itemize}
For completeness we give also the analytic expressions of the double-
and single-logarithmic corrections that contain logarithms
$\log{(\MW^2/\la^2)}$ and $\log(\MW^2/m_f^2)$, which involve the
photon mass $\la$ or masses of light charged fermions.\footnote{This
  kind of contributions includes also energy-dependent double
  logarithms of the type
  $\alpha\log{(\hat{s}/\MW^2)}\log{(\MW^2/\la^2)}$.}  These
contributions, denoted by $L^\elm$ and $l^\elm$ in the following, are
of pure electromagnetic origin and are not included into the numerical
studies.

The coefficients of the various logarithmic terms are expressed in
terms of the eigenvalues $I^{V^a}_\varphi$, or of the matrix
components $I^{V^a}_{\varphi\varphi'}$, of the generators\footnote{A
  detailed list of the gauge-group generators and of related
  quantities that are used in the following can be found in App.~B of
  \citere{Denner:2001jv}.}
\beq
I^A=-Q=-\frac{Y}{2}-T^3,\qquad 
I^Z=-\frac{\sw}{2\cw}Y+\frac{\cw}{\sw}T^3,\qquad
I^\pm=
\frac{T^1\pm\ri T^2}{\sqrt{2}\sw},
\eeq
where $\cw^2=1-\sw^2=\MW^2/\MZ^2$.

\subsection{Born matrix elements in the high-energy limit}

As input for the evaluation of the relative corrections \refeq{relco}
we need the Born matrix elements for the processes \refeq{duprocess}
and the $\SUtwo$-transformed Born amplitudes that we list in the
following, restricting ourselves to the non-suppressed helicities.
The corresponding amplitudes, are given in \refeq{longbornWN} and
\refeq{trabornWN} in high-energy approximation, \ie omitting
mass-suppressed terms. As we will see, this leads to very compact
analytical expressions for the relative corrections \refeq{relco}.
However, we recall that in the numerics instead of the high-energy
approximations \refeq{longbornWN} and \refeq{trabornWN} the
corresponding exact expressions have been used.  As noted in
\refse{se:WZRC}, the difference is at the per-mille level.

\newcommand{\Hdu}{H}%
\newcommand{\sN}{\tilde{N}}%
\begin{figure}
\unitlength 1cm 
\begin{center}
\begin{picture}(15,2.5)
\put(1.2,0){
\begin{picture}(120,80)
\Text(-.8,2.2)[r]{a)}
\ArrowLine(20,30)(0,60)
\ArrowLine(0,0)(20,30)
\Vertex(20,30){2}
\Photon(20,30)(80,30){2}{6}
\Vertex(80,30){2}
\DashLine(100,0)(80,30){2}
\DashLine(80,30)(100,60){2}
\Text(-0.1,0.1)[r]{$\Pu_\rL$}
\Text(-0.,2.)[r]{$\bar{\Pd}_\rL$}
\Text(3.5,0.1)[l]{$\chi$}
\Text(3.5,2.)[l]{$\phi^+$}
\Text(1.8,1.2)[b]{$\PWp$}
\end{picture}}
\put(6.9,0){
\begin{picture}(120,80)
\Text(-.4,2.2)[r]{b)}
\ArrowLine(40,45)(10,60)
\ArrowLine(40,15)(40,45)
\ArrowLine(10,0)(40,15)
\Vertex(40,15){2}
\Vertex(40,45){2}
\Photon(70,0)(40,15){2}{3}
\Photon(40,45)(70,60){-2}{3}
\Text(0.3,0.1)[r]{$\Pu_\rL$}
\Text(0.4,2.)[r]{$\bar{\Pd}_\rL$}
\Text(1.3,1.2)[r]{$\Pu_\rL$}
\Text(2.6,0.)[l]{$N_\rT$}
\Text(2.6,2.1)[l]{$\PW^+_\rT$}
\end{picture}}
\put(11.4,0){
\begin{picture}(120,80)
\ArrowLine(40,45)(10,60)
\ArrowLine(40,15)(40,45)
\ArrowLine(10,0)(40,15)
\Vertex(40,15){2}
\Vertex(40,45){2}
\Photon(70,0)(40,45){2}{5}
\Photon(70,60)(40,15){-2}{5}
\Text(0.3,0.1)[r]{$\Pu_\rL$}
\Text(0.4,2.)[r]{$\bar{\Pd}_\rL$}
\Text(1.3,1.2)[r]{$\Pd_\rL$}
\Text(2.6,0.)[l]{$N_\rT$}
\Text(2.6,2.1)[l]{$\PW^+_\rT$}
\end{picture}}
\end{picture}
\end{center}
\caption{Dominant lowest-order diagrams for
  $\bar{\Pd}_\rL\Pu_\rL\to\phi^+\chi$ and
  $\bar{\Pd}_\rL\Pu_\rL\to\PW^+_\rT N_\rT$}
\label{duborn}
\end{figure}
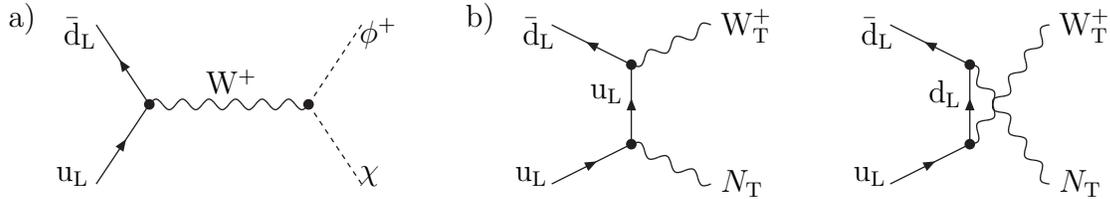%

For longitudinally polarized gauge bosons, we consider the Born matrix
elements involving the corresponding would-be Goldstone bosons.  These
are dominated by the $s$-channel exchange of gauge bosons (see
\reffi{duborn}a), and read
\beqar\label{longbornWN} 
\M_{\Born}^{\bar{\Pd}_\rL\Pu_\rL\rightarrow\phi^+\chi}&=&
\frac{-\ri e^2}{2\sqrt{2}\sw^2}\frac{A_s}{\hat{s}},\qquad
\M_{\Born}^{\bar{q}_\rL q_\rL\rightarrow H\chi}=e^2I^Z_{q_\rL}I^Z_{\chi H}
\frac{A_s}{\hat{s}},
\qquad
\M_{\Born}^{\bar{q}_\rL q_\rL\rightarrow \chi\chi}=0,
\nl
\M_{\Born}^{\bar{q}_\rL q_\rL\rightarrow \phi^+\phi^-}&=&
-e^2 \left(\frac{T^3_{q_\rL}}{2\sw^2}+\frac{Y_{q_\rL}}{4\cw^2}\right)
\frac{A_s}{\hat{s}},\qquad q_\rL = \Pu_\rL,\Pd_\rL.
\eeqar
The production of transverse  gauge bosons is dominated by the $t$- and 
$u$-channel contributions (see \reffi{duborn}b) and gives
\beqar
\label{trabornWN} 
\M_{\Born}^{\bar{\Pd}_\rL\Pu_\rL\rightarrow \PW_\rT^+ N_\rT }&=&
\frac{e^2}{\sqrt{2}\sw} \left(I^N_{\Pu_\rL}\frac{1}{\hat{t}}+I^N_{\Pd_\rL}\frac{1}{\hat{u}}\right)A_{t}, 
\nl
\M_{\Born}^{\bar{q}_\rL q_\rL\rightarrow N'_\rT N_\rT }&=&
e^2 I^{N'}_{q_\rL}I^N_{q_\rL} \left(\frac{1}{\hat{t}}+\frac{1}{\hat{u}}\right)A_t,\qquad
\M_{\Born}^{\bar{q}_\rL q_\rL\rightarrow \PW^+_\rT \PW^-_\rT}=
\frac{e^2}{2\sw^2}\frac{A_t}{\hat{r}}
,
\eeqar
where $\hat{r}=\hat{t},\hat{u}$ for $q=\Pd,\Pu$, respectively. In
order to determine the relative corrections, the explicit dependence
of the amplitudes $A_s$ and $A_t$, in \refeq{longbornWN} and
\refeq{trabornWN}, on the kinematics and on the helicities need not to
be specified.

\subsection{Leading soft--collinear corrections}
The angular-independent leading soft--collinear (LSC) corrections,
which are given in Eqs.~(3.6) and (3.7) of \citere{Denner:2001jv},
depend on the eigenvalues
\beq
\cew_\Phi=\frac{1+2\cw^2}{4\sw^2\cw^2},\qquad
\cew_{\Pu_\rL}=\cew_{\Pd_\rL}=\cew_{q_\rL}=\frac{\sw^2+27\cw^2}{36\sw^2\cw^2},\qquad
\cew_\PW=\frac{2}{\sw^2}
\eeq
of the electroweak Casimir operator $\cew$, on its components 
\beq
\cew_{AA} =2,\qquad  \cew_{AZ} = \cew_{ZA} =-2\frac{\cw}{\sw},\qquad  
\cew_{ZZ} =2\frac{\cw^2}{\sw^2}
\eeq
in the neutral gauge-boson sector, as well as on the squared \PZ-boson 
couplings
\beqar
(I^Z_{\bar{\Pd}_\rL})^2&=&\frac{(3\cw^2+\sw^2)^2}{36\sw^2\cw^2},\qquad
(I^Z_{\Pu_\rL})^2= \frac{(3\cw^2-\sw^2)^2}{36\sw^2\cw^2},\nl
(I^Z_{\PWm})^2&=&\frac{\cw^2}{\sw^2},\qquad
(I^Z_{\phi^-})^2= \frac{(\cw^2-\sw^2)^2}{4\sw^2\cw^2},\qquad
(I^Z_{\chi})^2= \frac{1}{4\sw^2\cw^2}.
\eeqar
For longitudinal and transverse gauge bosons we have
\beqar\label{SCWZcorr}
\de^{\SC}_{\bar{\Pd}_\rL\Pu_\rL \to \PW^+_{\rL} \PZ_{\rL}}&=&
\frac{\alpha}{4\pi}\biggl\{-\left[\cew_{q_\rL}+\cew_\Phi \right]\Los
\nl&&{}
+\sum_{\varphi=\bar{\Pd}_\rL,\Pu_\rL,\phi^-,\chi} 
(I^Z_\varphi)^2 \los\loZW
\biggr\}
\nl&&{}
-\frac{1}{2}\sum_{\varphi=\bar{\Pd}_\rL,\Pu_\rL,\phi^-}
Q_\varphi^2\Lemphi,
\eeqar
and
\newcommand{\factG}{G}
\beqar\label{duSCgen}
\de^{\SC}_{\bar{\Pd}_\rL\Pu_\rL \to \PW^+_{\rT} N_{\rT}}&=&
\frac{\alpha}{4\pi}\left\{-\frac{1}{2}\left[\sum_{\varphi=
\bar{\Pd}_\rL,\Pu_\rL,\PW^-} \cew_{\varphi}+ 
\sum_{N'}  \cew_{N'N} 
\frac{\M_{\Born}^{\bar{\Pd}_\rL\Pu_\rL\to\PW^+_{\rT} N'_{\rT}}}
{\M_{\Born}^{\bar{\Pd}_\rL\Pu_\rL \to\PW^+_{\rT} N_{\rT}}}\right]
\Los
\right.\nl&&\left.{}
+\sum_{\varphi=\bar{\Pd}_\rL,\Pu_\rL,\PW^-}(I^Z_\varphi)^2 \los\loZW
\right\} \nl&&{}
-\frac{1}{2} \sum_{\varphi=\bar{\Pd}_\rL,\Pu_\rL,\PW^-} Q_\varphi^2\Lemphi,
\eeqar
respectively, where the electromagnetic logarithms 
$L^\elm(\hat s,\la^2,M^2_\varphi)$ are given by 
\beq
L^\elm(\hat{s},\la^2,m^2_\varphi)
= \frac{\alpha}{4\pi}\left\{
2\los\loWla+
\log^2{\left(\frac{\MW^2}{\la^2}\right)}-
\log^2{\left(\frac{m^2_\varphi}{\la^2}\right)}
\right\}.
\eeq
For transverse final states, the non-diagonal components $\cew_{AZ}$
and $\cew_{ZA}$ of the electroweak Casimir operator require the
evaluation of the transformed matrix elements with $N'\neq N$.  Using
the high-energy approximation of the Born matrix elements
\refeq{trabornWN}, \refeq{duSCgen} can be written as
\beqar\label{duSCgen2}
\de^{\SC}_{\bar{\Pd}_\rL\Pu_\rL \to \PW^+_{\rT} N_{\rT}}&=&
\frac{\alpha}{4\pi}\left\{
-\left[\cew_{q_\rL}+\frac{1}{2}\cew_W\left(1+ \factG^N_- \right)\right]\Los
\right.\nl&&\left.{}+\sum_{\varphi=\bar{\Pd}_\rL,\Pu_\rL,\PW^-}(I^Z_\varphi)^2
\los\loZW \right\}\nl&&{}
-\frac{1}{2}\sum_{\varphi=\bar{\Pd}_\rL,\Pu_\rL,\PW^-} Q_\varphi^2\Lemphi,
\eeqar
with the angular-dependent 
functions
\beq\label{Gfact2}
\factG_\pm^A=\frac{F_\pm}{F_-+Y_{q_\rL}F_+},\qquad
\factG_\pm^Z=\frac{\cw^2F_\pm}{\cw^2F_--\sw^2 Y_{q_\rL}F_+},
\eeq
and
\beq
F_{\pm}=\left(\frac{1}{\hat{t}}\pm\frac{1}{\hat{u}}\right).
\eeq

\subsection{Subleading soft--collinear corrections}

The angular-dependent subleading soft--collinear ($\SS$) corrections
are obtained by applying the formula (3.12) of \citere{Denner:2001jv},
to the crossing symmetric process $\bar{\Pd}_\rL(p_{\bar{\Pd}})\,
\Pu_\rL(p_\Pu)\, \PW^-_{\la_\PW}(-p_\PW)\, N_{\la_N}(-p_N) \to 0$,
with $r_{12}=\hat{s}$, $ r_{13}=\hat{t}$, and $r_{14}=\hat{u}$. This
yields
\beqar\label{SSCcorrections}
\lefteqn{
\de^{\SS}_{\bar{\Pd}_\rL\Pu_\rL \to \PW^+_{\la} N_{\la}}=
\frac{\alpha}{4\pi}\sum_{V^a= A,Z}2\left[\los +\loWVa\right]}\quad&&\nl
&&\qquad\times I^{V^a}_{\PW^-_{\la}}\left[I^{V^a}_{\bar{d}_\rL}\lots+I^{V^a}_{u_\rL}\lous\right]\nl
&&
{}+\frac{\alpha}{4\pi}\left\{
2\los  \sum_{N'_{\la}}I^Z_{N'_{\la}N_{\la}} \M_{\Born}^{\bar{\Pd}_\rL\Pu_\rL\to\PW^+_{\la} N'_{\la} }
\left[I^Z_{\bar{d}_\rL}\lous+I^Z_{u_\rL}\lots\right]
\right.\nl&&\left.
{}-\frac{2}{\sqrt{2}\sw}\los
\left[
\left(\sum_{N'_{\la}}
I^+_{N'_{\la}} \M_{\Born}^{\bar{\Pu}_\rL\Pu_\rL \to N'_{\la} N_{\la}}
+I^+_{N_{\la}}\M_{\Born}^{\bar{\Pd}_\rL\Pd_\rL \to\PW^+_{\la} \PW^-_{\la}}
\right)\lots
\right.\right.\nl&&\left.\left.
{}-\left(\sum_{N'_{\la}}
I^+_{N'_{\la}} \M_{\Born}^{\bar{\Pd}_\rL\Pd_\rL\to N'_{\la} N_{\la}}
+I^+_{N_{\la}}\M_{\Born}^{\bar{\Pu}_\rL\Pu_\rL \to\PW^+_{\la} \PW^-_{\la}}
\right)\lous\right]\right\}
\left(\M_{\Born}^{\bar{\Pd}_\rL\Pu_\rL\to\PW^+_{\la} N_{\la}} \right)^{-1}
.\nln
\eeqar
In the cases $\la=\rL$ and $\la=\rT$, the sums run over
$N'_\rL=\chi,H$ and $N'_\rT=\PA,\PZ$, respectively, and
$I^+_{N_{\la}}$ are defined in (B.23) and (B.27) of \citere{Denner:2001jv}.  
Using the $\SUtwo$-transformed Born matrix elements given in
\refeq{longbornWN} and \refeq{trabornWN}, we obtain
\beqar\label{SSCcorr2}
\de^{\SS}_{\bar{\Pd}_\rL\Pu_\rL \to \PW^+_{\rL} \PZ_{\rL}}
&=&-\frac{\alpha}{2\pi}\frac{1}{\sw^2}\los \left[ \lous+\lots
-\frac{\sw^2}{\cw^2}Y_{q_\rL}\lotu
\right]\nl
&&{}+2\lemW \left[Q_\Pd \lots - Q_\Pu\lous \right],\nl
\de^{\SS}_{\bar{\Pd}_\rL\Pu_\rL \to \PW^+_{\rT} N_{\rT}}
&=&-\frac{\alpha}{2\pi}
\frac{1}{\sw^2}\los \left[
\lots+\lous+ \factG^N_+\lotu
\right]\nl
&&{}+2\lemW \left[Q_\Pd \lots - Q_\Pu\lous \right],
\eeqar
where 
\beq
l^\elm(M^2)={\alpha\over{4\pi}}\left [{1\over 2}\log\left ({\MW^2\over{M^2}}
\right )+\log\left ({\MW^2\over{\lambda^2}}\right )\right ],
\eeq 
and $G^N_+$ is given in \refeq{Gfact2}. 

\subsection{Single-logarithmic  corrections}
The single-logarithmic corrections consist of the contributions
$\de^\cc$ and $\de^\pre$ described in \refse{app:one-loopcorr}.  For
longitudinally polarized final states, according to Eqs.~(4.6) and
(4.33) in \citere{Denner:2001jv}, the corrections $\de^\cc$ read
\beqar\label{CcorrectionsL}
\de^\cc_{\bar{\Pd}_\rL\Pu_\rL \to\PW_\rL^+ \PZ_\rL }&=&\frac{\alpha}{4\pi} 
\left[\left(3 \cew_{q_\rL}+ 4\cew_\Phi\right)\los -\frac{3}
{2\sw^2}\frac{\Mt^2}{\MW^2}\losmt\right]
\nl &&
+\sum_{\varphi=\bar{\Pd}_\rL,\Pu_\rL,\PW^-}Q_\varphi^2\lemphi,\nln
\eeqar
and  the parameter renormalization yields
\beqar\label{RGWW}
\de^\pre_{\bar{\Pd}_\rL\Pu_\rL \to \PW_\rL^+ \PZ_\rL }
 &=&-\frac{\alpha}{4\pi}\bew_{W}\los +\Delta\alpha(\MW^2),
\eeqar
where $\bew_{W}=19/(6\sw^2) $ is the one-loop coefficient of the
$\SUtwo$ $\be$-function, and $\Delta\alpha(\MW^2)$ represents the
running of the electromagnetic coupling constant from the scale $0$ to
$\MW$.

If the final-state gauge bosons are transversely polarized then the
$\log{(\hat{s}/\MW^2)}$ contributions in $\de^\cc$ which are
associated to the final gauge bosons cancel the
$\log{(\hat{s}/\MW^2)}$ contributions originating from parameter
renormalization, and according to Eqs.~(4.6) and an analogue of (A.11)
in \citere{Denner:2001jv} one obtains
\beqar\label{eq:deCPRTT}
\lefteqn{\de^\cc_{\bar{\Pd}_\rL\Pu_\rL \to\PW_\rT^+ N_\rT }+
\de^\pre_{\bar{\Pd}_\rL\Pu_\rL \to \PW_\rT^+ N_\rT }
=}\quad&&\nl&=& 
\frac{3\alpha}{4\pi} \cew_{q_\rL}\los +\sum_{\varphi=\bar{\Pd}_\rL,\Pu_\rL,\PW^-}Q_\varphi^2\lemphi+\frac{1}{2}(1+\de_{N\PZ})\Delta\alpha(\MW^2),
\eeqar
where $\de_{N\PZ}$ represents the Kronecker symbol.

\end{appendix}

\end{document}